\documentclass[11pt,peerreview]{ieeeaccess}

\usepackage{cite}
\usepackage{amsmath,amssymb,amsfonts}
\usepackage{algorithmic}
\usepackage{graphicx}
\usepackage{textcomp}
\def\BibTeX{{\rm B\kern-.05em{\sc i\kern-.025em b}\kern-.08em
    T\kern-.1667em\lower.7ex\hbox{E}\kern-.125emX}}

\usepackage{adjustbox}
\usepackage{multirow}
\usepackage{booktabs,tabularx,caption}

\usepackage{eso-pic}
\usepackage{atbegshi,picture}
\usepackage[usestackEOL]{stackengine}

\usepackage{calligra}
\DeclareMathAlphabet{\mathcalligra}{T1}{calligra}{m}{n}

\DeclareMathAlphabet{\mathlcal}{U}{dutchcal}{m}{n}
\SetMathAlphabet{\mathlcal}{bold}{U}{dutchcal}{b}{n}

\newcommand{\propnumber}{} 

\newcommand{\thmnewnumber}{} 

\newcommand{\asmnumber}{} 
\newtheorem{assumption}{Assumtption\asmnumber}

\newcommand{\defnumber}{} 

\newcommand{\asmnumberC}{C.\hspace{-0.25em}} 

\newcommand{\exmnumber}{} 

\newcommand\Bigger[1]{\mathlarger{\mathlarger{#1}}}

\newcommand\abs[1]{\left| {#1} \right|}
\newcommand\close[1]{\left( {#1} \right)}
\newcommand\braket[1]{\left[ {#1} \right]}
\newcommand\curly[1]{\left \{{#1} \right\}}

\newcommand\myspace{\,\,\,}

\newcommand\fracBig[2]{\frac{\mathlarger{{#1}}}{\mathlarger{{#2}}}}

\newcommand\B[1]{\boldsymbol{{#1}}}

\newcommand\disturb{\xi}

\newcommand{\norm}[2]{\left\lVert#1\right\rVert_{#2}}
\newcommand{\normsq}[2]{\Big\lVert{#1}\Big\rVert_{#2}^{2}}

\newcommand\LeftCurly[2]{\scaleobj{#1}{\left\{\right.}{#2}}
\newcommand\RightCurly[2]{{#2}\scaleobj{#1}{\left.\right\}}}
\newcommand\LeftClose[2]{\scaleobj{#1}{\left(\right.}{#2}}
\newcommand\RightClose[2]{{#2}\scaleobj{#1}{\left.\right)}}
\newcommand\LeftBraket[2]{\scaleobj{#1}{\left[\right.}{#2}}
\newcommand\RightBraket[2]{{#2}\scaleobj{#1}{\left.\right]}}

\makeatletter
\def\Prob{\@ifnextchar[{\@with}{\@without}}
\def\@with[#1]#2{\myProbTwo{#2}{#1}}
\def\@without#1{\myProb{#1}}
\makeatother

\makeatletter
\def\ProbS{\@ifnextchar[{\@withS}{\@withoutS}}
\def\@withS[#1]#2{\myProbSTwo{#2}{#1}}
\def\@withoutS#1{\myProbS{#1}}
\makeatother

\makeatletter
\def\NonProb{\@ifnextchar[{\@withNon}{\@withoutNon}}
\def\@withNon[#1]#2{\myNonProbTwo{#2}{#1}}
\def\@withoutNon#1{\myNonProb{#1}}
\makeatother

\newcommand\myProb[1]{\mathrm{Pr}(\mathrm{y_{2}^*}\ge 0|\B{\mathrm{{#1}}={#1}_{g_i}})}

\newcommand\myProbTwo[2]{\mathrm{Pr}(\mathrm{y_{2}^*}\ge 0|\B{\mathrm{{#1}}={#1}_{g_i},\mathrm{{#2}}={#2}_i})}

\newcommand\myProbS[1]{\mathrm{Pr}(\mathrm{S}=1 |\B{\mathrm{{#1}}={#1}})}

\newcommand\myProbSTwo[2]{\mathrm{Pr}(\mathrm{S}=1|\B{\mathrm{{#1}}={#1}_{g_i},\mathrm{{#2}}={#2}_i})}

\newcommand\myNonProb[1]{\mathrm{Pr}(\mathrm{y_{2}^*}< 0|\B{\mathrm{{#1}}={#1}_{g_i}})}

\newcommand\myNonProbTwo[2]{\mathrm{Pr}(\mathrm{y_{2}^*}< 0|\B{\mathrm{{#1}}={#1}_{g},\mathrm{{#2}}={#2}_i})}

\usepackage{accents}

\usepackage{setspace}  
\usepackage{scalerel}

\usepackage{relsize}

\IEEEoverridecommandlockouts

\begin{document}
\history{This article has been accepted for publication in a future issue of this journal, but has not been fully edited. Content may change prior to final publication. Citation information: DOI 10.1109/ACCESS.2018.2888575, IEEE Access}
\doi{10.1109/ACCESS.2018.2888575}

\title{SEMIPARAMETRIC CORRECTION FOR ENDOGENOUS TRUNCATION BIAS WITH VOX POPULI BASED PARTICIPATION DECISION}
\author{\uppercase{Nir Billfeld}\authorrefmark{1},
\uppercase{Moshe Kim}\authorrefmark{2} %
}

\address[1]{University of Haifa, Haifa, Israel (e-mail: nbillfeld@staff.haifa.ac.il)}
\address[2]{University of Haifa, Haifa, Israel (e-mail: kim@econ.haifa.ac.il)}


\markboth
{This article has been accepted for publication in a future issue of this journal, but has not been fully edited.}
{This article has been accepted for publication in a future issue of this journal, but has not been fully edited.}

\corresp{Corresponding author: Moshe kim (e-mail: kim@econ.haifa.ac.il).}

\begin{abstract}
We synthesize the knowledge present in various scientific disciplines for the development of semiparametric endogenous truncation-proof algorithm, correcting for truncation bias due to endogenous self-selection. This synthesis enriches the algorithm's accuracy, efficiency and applicability. Improving upon the covariate shift assumption, data are intrinsically affected and largely generated by their own behavior (cognition). Refining the concept of Vox Populi (Wisdom of Crowd) allows data points to sort themselves out depending on their estimated latent reference group opinion space. Monte Carlo simulations, based on 2,000,000 different distribution functions, practically generating 100 million realizations, attest to a very high accuracy of our model. 
\end{abstract}

\begin{keywords}
Expectation-Maximization, Fourier-based Sieve estimator, latent reference groups,  Local Covariate shift, Monte Carlo simulation, Opinion space, SCAD, Selectivity bias correction, Semiparametric, Vox Populi, Wisdom of crowds
\end{keywords}

\titlepgskip=-15pt

\maketitle

\IEEEpeerreviewmaketitle 
\section{Introduction}
An important fact, but one that is largely overlooked or taken-for-granted, is that researchers hardly ever have access to the entire data distribution pertaining to their specific research and rely, instead, on a truncated form of such data. The truncated data employed probably has different characteristics than the latent non-truncated full distribution and may result in biased parameter estimates generated by the specific investigated models. The problem is further aggravated when truncation is endogenously propagated by various decision units, or observations. Examples of a straight-forward endogenous truncation emerge from aspects of some type of discouragement. For instance, in labor markets, long-term unemployed persons are often discouraged workers who are afraid that they will not find employment and therefore  do not seek employment;  hence, they will be absent from reported unemployment rates. Another, gender-related, labor market example is discouraged women: it has been shown that most women do not apply for jobs that require a high degree of aggressiveness; hence we have discouraged women who fail to participate in specific sectors of the economy, which affects female labor supply. In financial markets, discouraged borrowers, such as some small and medium size enterprises, do not apply for loans and result in biased modeling of default probabilities, which hampers optimal credit allocation. Endogenous truncation also is involved in the measurement of social problems, such as the crime or divorce rates, because measurement may represent a latent rate of reporting, rather than the variables of interest. Similarly, endogenous truncation severely impacts the measurement of important economic indices, such as economic growth, productivity, income distribution, and welfare.

The notion of truncation is different than the concept of the known selection bias. In the known selection bias, information on data (observations) has been censored but still observable or, alternatively, information regarding the counterfactual (e.g., the rejected rather than the discouraged borrower) has been censored but still observable. Selection bias under censoring has already been remedied by Heckman's seminal contribution \cite{heckman1979sample}. Under Heckman's model, the selection process is entirely observed and selectivity bias can be alleviated. Under \textit{endogenous truncation}, however, the \textit{selection rule} is completely unobserved and no information is available concerning the truncated observations. Thus, statistical biases are myriad and interwoven to the extent that researchers may not even be able to assess their magnitude and direction, and the problem becomes  extremely challenging. Given the potential severity of the aforementioned problem, it is surprising that the endogenous truncation problem has attracted hardly any scientific investigation, assessment, or suggestions for proper remedies. 

In the few existing interests in the literature, the identification of the semiparametric truncated sample selection model is achieved by  observing the selection variable (which is modeled as continuous), while imposing different restrictions on the disturbances \cite{powell1994estimation,honore1997estimation}, or by utilizing information regarding some of the non-participants' characteristics \cite{khan2007weighted}. Both of these studies rely on available data regarding either the covariates' joint distribution function or the selection variable, implying that the variable  is not treated as a latent binary response variable (unlike the approach taken in the present paper). Further, these studies model the selection rule of each datum as a function of its \textit{observed} characteristics. Yet, the selection rule might be affected by unobserved (truncated) characteristics, as well. Ignoring these characteristics may lead to misspecification of the selection equation, potentially biasing the estimates. Additionally, the estimation and identification of semiparametric truncated sample selection models with a latent binary selection variable are known to be difficult, due to the absence of observed variation in exactly this selection variable. The various estimation procedures that utilize a continuous selection variable to alleviate this difficulty use different kernel estimators \cite{lee1993quadratic}. The closest approach to the proposed methodology, dealing with a latent binary selection variable, is \cite{ichimura1993semiparametric}, which also employs a kernel to estimate the bias term in the substantive equation \cite{ichimura1991semiparametric,ichimura1993semiparametric}. However, the resulting estimates can still be biased, as  the kernel estimator's accuracy depends on selecting the optimal bandwidth, which is hard to find in the semiparametric context \cite{lewbel2007simple}. 

An additional, important weakness of the existing literature dealing with endogenous truncation problems is the assumption of similar behavior on the part of the truncated and non-truncated distributions \cite{ichimura1993semiparametric}, an assumption which is referred to as a population regression, in the econometric literature \cite{heckman1979sample}, and a \textit{covariate shift}, in the computer science literature \cite{gretton2009covariate}. The various truncated sample selection models treat the data as if they all consist of a single, homogeneous, monolithic cohort sharing identical actions, such that the selection rule of each datum is not affected by the participation decisions of other members. This restrictive assumption, however, can introduce selection bias by itself. In fact, as \cite{manski2010consensus} describes it: ``If agents knew the state of nature, they would make the same decision. However, they may have different beliefs or may use different decision criteria to cope with their incomplete knowledge. Hence, they may use different actions even though they share the same objective'' (p.187). 

Taking into consideration that we are unable to observe the selection variable, we propose an estimation procedure. In order to rectify the aforementioned potential bias and to improve upon the \textit{covariate shift} assumption, which is frequently used in machine learning, the data in our model are treated as a mixture of sub-populations, each characterized by its own action regarding the participation decision. Thus, we build on the \textit{vox populi} concept \cite{galton1907vox} or, in its modern term, ``The Wisdom of Crowds'' \cite{surowiecki2005wisdom}, as the basis by which data points  ``sort themselves'' in the truncation process. As such, each data point's  ``decision'' to allow itself to be truncated  from the original distribution is an important building block that generates our offered algorithm.

The \textit{vox populi} concept relies on the idea that aggregates of opinions measuring the central tendency will be more accurate than individual opinions \cite{davis2014crowd}. \cite{budescu2005confidence} suggests that an aggregate  of multiple sources maximizes the amount of information available and reduces the potential impact of unreliable information sources. The implication is that  the combination of the various sources leads to error cancellation.

Further, we refine the concept of ``the wisdom of crowds''  to be a non monolithic concept and apply it to truncation. Each observation ``decides'' whether to allow itself to be truncated depending on its reference group's (rather than on the entire crowd's) decision  opinion space average forecast. This, in turn,  is inspired by the similarity-based classification in machine learning and Cybernetics \cite{chen2009similarity,hummel1996statistical} and management science models of decision making \cite{budescu2014identifying}. This enables the various opinion spaces, generated by the various reference groups, to provide expert opinion with rather superior average forecast, by eliminating poorly-performing individuals from the crowd \cite{prelec2017solution,budescu2014identifying}. Such treatment is also inspired by economic theories of ethnic capital \cite{Borjas1992} and informational cascades \cite{bikhchandani1992theory},  highlighting the fact that individual characteristics depend on the average characteristics of the group to which they belong.  Recently, we have witnessed an upsurge of interest in the relationship between culture and genetic diversity \cite{desmet2017culture}, through the process of endogenous group selection \cite{ashraf2013Genetic}. 

Building upon this insight, we model the number and type of reference groups to be endogenously determined, rather than arbitrarily imposed. A Latent Class Analysis (LCA) \cite{clogg1984latent} is used to estimate the latent characteristics (type) of the various reference groups. This is implemented by integrating Machine Learning concepts and providing a \textit{Fourier}-based Sieve semiparametric estimator, which is distribution-free. Our estimator uses a penalized non-linear regression \cite{schuurmans2002metric}, an important characteristic emphasizing the generality and applicability of the offered methodology. The \textit{Fourier series} is a functional of the \textit{Orthonormal} polynomials sequence family, which allows for efficient estimation of functions with non-smoothness, discontinuities in derivatives, sharp spikes and discontinuities in the function itself. Thus, it is useful in nonparametric regression for approximating a much broader class of functions \cite{ogden2012essential} than the kernel approach. 

The most attractive feature of our proposed estimator is that it intrinsically prevents potential multicollinearity problems. Even though the multicollinearity might arise in certain circumstances, we can prevent it. For example, multicollinearity might arise if we extend the model by incorporating an endogenous covariate in the substantive equation and estimate sequentially a system of partially linear equations. The first equation is the endogenous covariate regression, which linearly depends on the selection bias term, while the second equation is the substantive equation (of interest), which linearly depends both on the endogenous covariate, as well as on a similar selection bias term. These two selection bias terms depend on the same covariate vector and thus they might be correlated. However, this problem is alleviated by the fact that each selection bias term is approximated by a different \textit{orthonormal polynomial sequence} (a different number of mutually orthogonal basis functions), which implies, by definition of orthonormality, that these two approximated functions cannot be perfectly multicollinear. This result is required for identification.\footnote{The identification can be achieved due to the fact that some of the mutually orthogonal basis functions (covariates) are not common to both series expansions. These non-common covariates play the role of an exclusion restriction which is commonly used to assure identification.} We note that the classical kernel estimator does not possess this advantageous orthonormality feature and consequently may produce biased estimates due to cross- equation correlation. 

Another aspect that our estimator must consider is the optimal number of groups. In order to find the optimal number of groups that best fits the data generation process, we perform variable selection (also referred to as "sparse regression" \cite{friedman2012fast,yang2016sparse}) by employing the smoothly clipped absolute deviation (SCAD) penalty function.\footnote{The SCAD penalty function is superior to the often employed least absolute shrinkage and selection operator (LASSO), because it is general and nests the LASSO as a special case.} We develop a generic non-linear penalized regression estimation method, in the sense that it can easily be extended to enable a wide collection of penalty functions to be estimated. The novelty of our modeling lies in the integration and synthesis of knowledge present in various scientific disciplines, such as: (i) computer science (pattern recognition, unsupervised machine learning,\footnote{For a constructive overview of the field of unsupervised learning, see \cite{ghahramani2004unsupervised}.} artificial intelligence and self-organizing maps in neural networks); (ii) electrical engineering (signal extraction); (iii) economics and; (iv) management for the creation of new algorithms correcting for truncation bias, due to the endogenous self-selection of observations into a sample. This integration enriches the algorithms' accuracy, efficiency and applicability and hopefully can be of use in economics and many other disciplines. 

We offer a three-stage procedure to correct for the endogenous truncation bias: in the first stage, latent classes analysis is employed based on results from an auxiliary survey data, consisting of experts' (binary) opinions, as well as of their observed group characteristics, to recover the unobserved latent reference groups.\footnote{For example, the Small Business Credit Survey administered by \cite{FederalNY} is an annual survey of firms with fewer than 500 employees reporting on financing needs and choices and borrowing experiences. Based on the small business credit survey 2016, a total of $17\%$ of the non-applicants are discouraged borrowers.} A given expert's opinion captures his belief regarding the expected participation decision in his reference group.\footnote{Since the opinion is binary, each expert is asked what is the most likely decision for a random member belonging to her reference group being a participant or a non-participant.} In the second stage, each participant share is obtained by averaging the members' opinions belonging to the specific reference group.\footnote{The average of opinions belonging to a particular reference group reflects  a refined version of the (monolithic) wisdom of crowds.} In the third stage, a semiparametric truncated sample selection model is estimated, consisting of a selection equation and a substantive equation. The estimated participants' share in the group, conditional on the reference group's observed and unobserved characteristics, is included in the selection equation as an additional covariate.  We run Monte Carlo simulations in order to examine our estimator's performance in the presence of a truncated sample selection model. 
Further, for sake of generality of the offered estimator, we subject it to various distributions in which the disturbances are neither jointly nor marginally normally distributed. These disturbances are constructed as realizations of non-symmetric and non-unimodal distribution functions.\footnote{Unlike the practice in some other studies applying only normally distributed disturbances.}

The rest of the paper is organized as follows: Section \ref{Section:Model} introduces the model consisting of substantive and selection equations; Section \ref{sec:Methodology} deals with model estimation; Section \ref{Unknown:Latent:Groups} recovers the number of reference groups; Section \ref{Sec:Simulations} examines our truncated selection model's performance, employing Monte Carlo simulations; and Section \ref{sec:Summary} concludes by summarizing the main findings, as well as our estimator's performance.

Next we present our suggested methodology for estimation of a truncated endogenous sample selection model in the presence of a reflection problem, when the entire data consist of participants only. 

\section{The model}\label{Section:Model}

\subsection {Binary choice selection equation}\label{Section:Binary:Choice:Selection}

We describe the participation choice of each individual observation $i\in\curly{1,...,N}$, as a function of its reference group's participation decision which is captured by the participants' share in the particular reference group. A model in which an individual's decision is affected by the average decision made by all its group members is referred to as the ``reflection problem'' \cite{manski1993dynamic},\footnote{The reflection problem arises \cite{manski1993dynamic} ``when a researcher observing the distribution of behavior in a population tries to infer whether the average behavior in some group influences the behavior of the individuals that comprise the group. The term reflection is appropriate, because the problem is similar to that of interpreting the almost simultaneous movements of a person and his reflection in a mirror. Does the mirror image cause the person's movements or reflect them?'' (p. 532)} or Manski's notion of role models/emulation \cite{manski1993identification}. These concepts may touch on an earlier idea of ``ethnic capital'' \cite{Borjas1992}, showing individual characteristics to be dependent on the average characteristics of the group they belong to, and a tendency to follow the decision of others \cite{bikhchandani1992theory}.

Let the number of reference groups (unknown to the researcher) be denoted by $G$. The individual choices given a membership in reference group $g\in\curly{1,...,G}$ are coded by $\omega_{i,g}\in\curly{0,1}$ and are defined as:
\begin{IEEEeqnarray}{lCr}\label{Individual:Decision}
\omega_{i,g}=
\begin{cases}
1 &  \text{a participant}   \\
0 &  \text{a non-participant}
\end{cases}
\end{IEEEeqnarray}  

These individual choices are determined by two sets of factors. The first set consists of the observed group-level characteristics $\B{x}_{g}\in\mathbb{R}^{L_{\B{x}}}$\footnote{The notation $\mathbb{R}^{L_{\B{x}}}$ stands for a vector of size $L_{\B{x}}\times 1$.} and the unobserved group-level characteristics, captured by a latent categorical variable $\psi_{i,g}$ of $\tilde{G}$ different outcomes, where $\tilde{G}$ is not arbitrarily imposed (as will be depicted in section \ref{Sec:Latent:Classes} to follow).\footnote{We allow for (but do not require) a dependence between the observed and unobserved group's characteristics, determined by some unknown joint distribution function (as depicted in section \ref{Sec:Latent:Classes} to follow).} The second set consists of the observed individual-level characteristics $\B{z_i}\in\mathbb{R}^{L_{\B{z}}}$  and an individual random disturbance  $\disturb_{2i}$.\footnote{Each reference group $g$ is a unique combination of observed  and unobserved characteristics ($\B{x}_g$ and $\psi_{i,g}$, respectively) which are common to all of the $g$'th reference group's members. However, the presence of unobserved characteristics $\psi_{i,g}$, implies that in order to assign observations into reference groups, $\psi_{i,g}$ is required to be estimated (as will be discussed in section \ref{Sec:Opinions:Experts} to follow).}

These factors are assumed to produce payoffs for the possible participation choices, $u_{i,g}(1)$ and $u_{i,g}(0)$, the utility of participation and non-participation, respectively.
The difference between these payoffs is additive in the various factors. A participation choice is made when the following difference is positive
\cite{brock2007identification}:\footnote{$T$ is defined everywhere in the manuscript as the transpose operator.}
\begin{IEEEeqnarray}{lcr}\label{Individual:Utility}
u_{i,g}(1)-u_{i,g}(0) = \alpha + \B{z}_i^T\B{\eta}\\  \hspace{2em}+\beta   m^{\mathlcal{e}}(\B{x}_g,\psi_{i,g})+\close{\B{x}_{g}^{c}}^T\B{\delta}-\disturb_{2i}& \nonumber
\end{IEEEeqnarray}
where $\B{x}_{g}^c$ is a subset of $\B{x}_{g}$ consisting of contextual factors,\footnote{ This decomposition is intended to satisfy the exclusion restriction in \eqref{Individual:Utility} for the sake of identification of the $\beta$ and $\delta$ parameters.}$^,$\footnote{A contextual effect exists whenever the propensity of a person to behave in some way varies with the characteristics of the reference group members.} $m^{\mathlcal{e}}(\B{x}_g,\psi_{i,g})$ is the expectation (forecast) of individual $i$ with reference group's characteristics $\B{x}_{g}$ and $\psi_{i,g}$ regarding the participants' share in his group, and the super-script $\mathlcal{e}$ represents expectation (forecast).\footnote{The subjective belief (forecast) is a mapping from group's (observed and unobserved) characteristics to a scalar representing a participation probability (participants' share).}

It is worth noting that the difference $u_i(1)-u_i(0)$ in \eqref{Individual:Utility} is positive iff the following inequality holds:\footnote{Instead of employing merely the average participation decision $m^{\mathlcal{e}}(\B{x}_g,\psi_{i,g})$, an interesting extension of this model would be to allow for each datum to be affected by a vector of moments (various dispersion measures) obtained from the survey.} 
\begin{IEEEeqnarray}{lCr}
\disturb_{2i}>\alpha + \B{z}_i^T\B{\eta} + \beta m^{\mathlcal{e}}(\B{x}_g,\psi_{i,g})+\close{\B{x}_{g}^{c}}^T\B{\delta}
\end{IEEEeqnarray}
which implies that the conditional participation probability given the reference group and individual level characteristics $\curly{\B{z_i},\B{x}_g,\psi_{i,g}}$  with $(\B{x}_{g}^c\subset \B{x}_g)$ is:
\begin{IEEEeqnarray}{lCr}\label{Bernoulli::Probability}
\mathrm{Pr}(\mathlarger{\omega_{i,g}}=1|\B{z_i},\B{x}_g,\psi_{i,g})=1-F_{\mathrm{\disturb_2}}(\alpha + \B{z}_i^T\B{\eta}\\  \hspace{2em} + \beta m^{\mathlcal{e}}(\B{x}_g,\psi_{i,g})+\close{\B{x}_{g}^{c}}^T\B{\delta})&\nonumber
\end{IEEEeqnarray}
where $F_{\mathrm{\disturb_2}}$ stands for the distribution function of the random disturbance $\disturb_{2i}$, which is unknown to the researcher, $\mathlarger{\mathlarger{\omega_{i,g}}}$ is a random variable that is conditionally Bernoulli-distributed, given the individual-level and group-level covariates,  while $\omega_{i,g}$ depicted in \eqref{Individual:Decision} stands for its realization.

Each individual is small relative to the population \cite{blume2010identification}. Using \eqref{Bernoulli::Probability}, the following condition is obtained:
\begin{IEEEeqnarray}{lCr}\label{Equilibrium:Condition}
m(\B{x}_g,\psi_{i,g})=\int \mathbb{E}[\mathlarger{\omega_{i,g}}|\B{\mathrm{z}},\B{x}_g,\psi_{i,g}]dF_{\B{\mathrm{z}}|\B{x}_{g}}\\=\int (1-F_{\mathrm{\disturb_2}}(\alpha + \B{\mathrm{z}}^T\B{\eta} + \beta m^{\mathlcal{e}}(\B{x}_g,\psi_{i,g})\nonumber\\   \hspace{2em}\nonumber\\+\close{\B{x}_{g}^{c}}^T\B{\delta})dF_{\B{\mathrm{z}}|\B{x}_{g}}& \nonumber
\end{IEEEeqnarray}
where $m(\B{x}_g,\psi_{i,g})$ is the actual participants' share, given a membership in a reference group characterized by observed and unobserved characteristics $\B{x}_g$ and $\psi_{i,g}$, respectively; $F_{\B{\mathrm{z}}|\B{x_{g}}}$ is the conditional distribution function of $\B{\mathrm{z}}$ (given $\B{x}_{g}$) which is unknown to the researcher.

We next present the theoretical model equations.

\subsection{The sample selection model}
The underlying model consists of two equations in which the latent (population) dependent variables $y_{1i,g}^*$ and $y_{i2,g}^*$ are defined as follows:
\begin{IEEEeqnarray}{lCr}\label{Latent:substantive:equation}
y_{1i,g}^*=\B{w}_i^T\B{\theta}+\disturb_{1i} \hspace{1em}\text{the susbtantive equation},
\end{IEEEeqnarray}
$\hspace{3em}$ and
\begin{IEEEeqnarray}{lCr}\label{IndividualParticipation:Decision}
y_{2i,g}^*=\alpha+\beta m^{\mathlcal{e}}(\B{x}_g,\psi_{i,g})+\B{z}_i^T\B{\eta}\\  \hspace{2em}+\close{\B{x}_{g}^{c}}^T\B{\delta}+\tilde{\disturb}_{2i}\hspace{1em}\text{the selection equation}\nonumber,
\end{IEEEeqnarray}
where $\B{w_i}\in\mathbb{R}^{L_w}$ and $\B{\theta}$ denote the substantive equation's covariate vector and a $L_w\times 1$ parameter vector, respectively. The substantive equation's random disturbance is $\disturb_{1i}$, and the selection equation's disturbance satisfies $\tilde{\disturb}_{2i}\equiv -\disturb_{2i}$.\footnote{Using the definition in \eqref{Individual:Utility}, $y_{2i,g}^*$ is the difference between the participation and non-participation utilities, which includes a random disturbance $\disturb_{2i}$ followed by a minus sign.} $\text{ }$The random disturbances $\disturb_{1i}$ and $\disturb_{2i}$, with their respective marginal distribution functions $F_{\disturb_1}$ and $F_{\disturb_2}$, are jointly distributed. Their joint distribution function is $F_{\disturb_1, \disturb_2}$. The model is semiparametric as neither the marginals nor the joint distribution function are required to be specified by the researcher.   $y_{ji,g}^{*}$ denotes a realization of the latent random variable $\mathrm{y}_j^*$  for $j=1,2$.\footnote{Asterisk implies a latent (population) variable.} The group-level characteristics $\B{x}_g$ and $\psi_{i,g}\in\curly{1,...,\tilde{G}}$ constitute the $i$'th observation's specific reference group; $m^{\mathlcal{e}}(\B{x}_g,\psi_{i,g})$ is the latent participants' share given a membership in latent reference group $(\B{x}_g,\psi_{i,g})$. 
$\beta$ captures the endogenous effect\footnote{The presence of an endogenous effect implies that the propensity of a person to behave in some way varies with the behavior of the reference group \cite{manski2000economic}. } and $\delta$ the contextual
effect.

In the truncated sample the $i$'th observation in group $g$ is denoted by the sequence $\curly{y_{1i,g},\B{x}_{g}^T,\B{w}_i^T,\B{z}_i^T}$, where $y_{1i,g}$ is defined as:
\begin{IEEEeqnarray}{lCr}\label{TruncatedEquation}
y_{1i,g}=\begin{cases} y_{1i,g}^*, & \mbox{if selected},\hspace{1em} y_{2i,g}^*\ge 0 \\ \text{Unobserved}, & \mbox{if not selected},\hspace{1em}y_{2i,g}^*< 0  \end{cases}
\end{IEEEeqnarray}

A binary random variable indicating participation is denoted by $\mathrm{S}^*$ defined as:
\begin{IEEEeqnarray}{lCr}\label{SelectionDummy}
\mathrm{S}^*=\begin{cases} 1, & \mbox{if selected},\hspace{1em} \mathrm{y}_{2}^*\ge 0 \\ 0, & \mbox{if not selected},\hspace{1em}\mathrm{y}_{2}^*< 0  \end{cases}.
\end{IEEEeqnarray}

However, $\mathrm{S}^*$ in \eqref{SelectionDummy} is unobserved, and only $\mathrm{S}$ is observed:
\begin{IEEEeqnarray}{lCr}
\mathrm{S}=\begin{cases} 1, & \mbox{if selected},\hspace{1em} \mathrm{y}_{2}^*\ge 0 \\ \text{Unobserved}, & \mbox{if not selected},\hspace{1em}\mathrm{y}_{2}^*< 0  \end{cases}.	
\end{IEEEeqnarray}

Let $n<N$ denote the number of observations in the truncated data set. The participants' share $m^{\mathlcal{e}}(\B{x}_g,\psi_{i,g})$ is a forecast of the actual participants' share $m(\B{x}_g,\psi_{i,g})$, and they are interrelated through \eqref{Equilibrium:Condition}.
\newcommand\prY[1]{\pi_{{_{#1}}}}	

Next we discuss the model estimation.
\section{Model Estimation}\label{sec:Methodology}
In this section, we propose an estimation procedure for a truncated selection model, consisting of a substantive equation and a selection equation. 

The estimation is a three steps sequential procedure: (i) A Latent Classes Analysis to estimate the reference groups' unobserved characteristics, as will be discussed in section \ref{Sec:Latent:Classes}; (ii) Evaluation of the participation probability in each reference group, controlling for its unobserved characteristics, by utilizing experts' opinions; and (iii) Estimating a partially linear index model using Sieve (series) estimator for the non linear component, which is referred to as the ``bias term''.

Next we discuss the main idea behind the assignment of each observation into latent classes, utilizing a survey data set consisting of experts' opinions and group-level covariates (a combination of continuous and categorical variables).

\newcommand\SizeSym[1]{\scaleobj{0.8}{\B{\mathlcal{{#1}}}}^E}
\newcommand\SizeSymGen[1]{\scaleobj{0.8}{\B{\mathlcal{{#1}}}}}
\newcommand\SizeKSym{\scaleobj{0.8}{\B{\mathlcal{K}}}}		
\subsection{Identification of latent reference groups}\label{Sec:Latent:Classes}
Latent Class Analysis (LCA) is a statistical method for matching a set of manifest (observed) variables to a set of latent variables referred to as classes \cite{goodman1974exploratory,clogg1984latent,greene2003latent}. A specific realization of the manifest variables is referred to as a ``response pattern''. Let $\mathfrak{Y}$ denote the set of response patterns consisting of all possible realizations of a $J\times 1$ categorical variable vector, defined as:
\begin{IEEEeqnarray}{lCr}\label{Manifest:variables}
\mathfrak{Y}=\curly{\braket{\mathlcal{y}_{_1},...,\mathlcal{y}_{_J}}^T\Big|\mathlcal{y}_{_j}\in\curly{1,...,K_j},\hspace{1em}j=1,...,J}\hspace{2em}
\end{IEEEeqnarray}
where the number of outcomes in the $j$'th categorical variable is $K_j$.

The role of the realizations of manifest variables in \eqref{Manifest:variables} is for identification purposes, by means of classifying observations into their most likely latent class utilizing recruitment probabilities. A recruitment probability is the probability that a specific response pattern $\scaleobj{0.8}{\B{\mathlcal{Y}}}\in\mathfrak{Y}$ will be observed for a randomly selected member of a given latent class.\footnote{The response pattern of the $i$'th observation is its set of responses to all the manifest variables. These responses are conditionally independent of each other in a given class.} The a posteriori probability of being a member in a given class is obtained by using Bayes' theorem as a function of the estimated recruitment probabilities and the estimated prevalence of each latent class (the class membership prior probability).
Each observation is assigned to the latent class that has the highest a posteriori probability. 

The number of latent classes (labels), $\tilde{G}$, is recovered by the model rather than arbitrarily imposed. The latent classes analysis is employed repeatedly for a given specific number of latent classes  $\mathlcal{k}\in\curly{2,...,\tilde{G}_{\max}}$.  $\tilde{G}_{\max}$ is the largest possible number of classes and is in the spirit of the Bayesian Information Criterion (BIC) \cite{schwarz1978estimating},   reported to perform well by finding the correct number of components in the mixture \cite{roeder1997practical}. Other authors suggest using
Bayesian-based graphical techniques to aid in deciding on the number of classes
\cite{garrett2000latent}. We depart from the aforementioned literature in that we apply the BIC criterion directly to the substantive equation, in order to find the best model specification under truncation by using SCAD (section \ref{Unknown:Latent:Groups}). To achieve this goal,  we employ a penalized non-linear regression model, using the SCAD penalty function, to select the best solution obtained from the latent classes analysis. 

We distinguish between two cases: (i) the class membership prior probabilities varies among observations, as these probabilities are determined by a covariates set; (ii) the class membership prior probabilities are constant across observations and there is no dependence on covariates. In the former, a parametric multinomial response model, such as the multinomial logistic regression, is employed to estimate the prior class membership given the covariates.\footnote{The justification for a parametric model is to reduce the complexity of calculations.} For the latter, one only needs to estimate $\mathlcal{k}-1$ class membership proportions (given $\mathlcal{k}$ classes) that characterize the entire data.\footnote{Without loss of generality, these unknown proportions can be estimated, nonparametrically, by a logistic multinomial response model characterized by a unique intercept per class. This is a nonparametric estimation procedure, due to the absence of covariates.}
As we focus on the endogenous determination of class membership, covariates are involved in the estimation of  the prior class membership probabilities. The model parameters that are required for the estimation of the labels' a posteriori distribution (in section \ref{Posterior:Classes} to follow) are: (i) the parameters which affect the class membership prior probability (in section \ref{Prior:classes} to follow); and (ii) the parameters which affect the response pattern given the class membership (the conditional response probabilities in section \ref{Recruitment:Classes} to follow). 

\subsection{The latent classes estimation}
In this section, we introduce an estimation procedure to recover the outcomes of the sequence $\curly{\psi_{i,g}}_{g=1}^G$, which are the reference groups' latent characteristics. Each outcome has its own label, and
the labels are estimated by employing latent classes analysis \cite{clogg1984latent,goodman1974exploratory}, a procedure to estimate their labels' a posteriori probability density function. Once this posterior function is estimated, the sequence of fitted labels are the arguments that maximize ($\arg \max$) the estimated posterior distribution function. 

	Suppose that the population consists of $\mathlcal{k}$ latent classes, such that the class membership of each observation $i=1,...N$ is denoted by an unobserved categorical variable $\psi_{i,g}$ with $\mathlcal{k}$ possible outcomes. We treat each observation as a random realization of the conditional labels' distribution function, given its groups' observed covariates $\B{x}_{g}$. This methodology is based on the non-random assignment into classes (heterogeneous class membership prior probabilities) introduced by \cite{clogg1984latent}.

Next we present the prior class membership probabilities under non-random assignment.
\subsubsection{The prior distribution function of the labels}\label{Prior:classes}
Let $\Psi_{\mathlcal{k}}$ be a categorical random variable  of $\mathlcal{k}$ potential outcomes. We denote the prior probability 
of belonging to label $t$, given the group's observed characteristics $\B{x}_g$ by $\lambda_{t|\B{x}_g}$, satisfying $\sum_{t=1}^{\mathlcal{k}}\lambda_{t|\B{x}_g}=1$ $\forall g\in\curly{1,...,G}$ and
defined as: 
\begin{IEEEeqnarray}{lCr}\label{Prior:Probability:Function}
\lambda_{t|\B{x}_g}\equiv\mathrm{Pr}(\Psi_{\mathlcal{k}}=t|\B{\mathrm{x}}=\B{x}_g)\\  \hspace{2em}=\frac{\exp\close{\Bigger{\varsigma}_{_{0,t}}+\B{x}_g^T\B{\mu}_t}}{1+\sum_{v=1}^{\tilde{G}-1}\exp\close{\Bigger{\varsigma}_{_{0,v}}+\B{x}_g^T\B{\mu}_{v}}},\hspace{1em} t=1,...,\mathlcal{k}\nonumber
\end{IEEEeqnarray}
where $\B{x}\in\mathbb{R}^{L_{\B{\mathrm{x}}}}$ is a group-level covariates vector, $\Bigger{\varsigma}_{_{0,1}},...,\Bigger{\varsigma}_{_{0,\mathlcal{k}}}$ are intercepts and $\B{\mu}_t\in\mathbb{R}^{L_{\B{\mathrm{x}}}}$ for $t=1,...,\mathlcal{k}-1$ are parameter vectors.\footnote{Although this function can be formulated nonparametrically, we have opted for the present multinomial logistic formulation for computational simplification. Latent classes analysis involves an iterative estimation procedure, and thus each iteration requires a different optimal bandwidth. Since we estimate 10,000 different data sets, the number of bandwidths to be computed would requires 10,000 times the number of iterations. Computationally, this is extremely cumbersome.}

Suppose, also, that conditional on $\B{x}_g$, $\Psi_{\mathlcal{k}}$ is jointly distributed with a vector of $J$ categorical variables (the group's manifest variables) $\B{\mathcal{Y}_i}=\braket{\mathcal{Y}_{1i},...,\mathcal{Y}_{Ji}}^T$, which is referred to as \textit{the vector of responses} and its realization is denoted by $\scaleobj{0.8}{\B{\mathlcal{Y}}}_i=\braket{\mathlcal{y}_{1i},...,\mathlcal{y}_{Ji}}^T\in\mathfrak{Y}$. The $j$'th observed categorical variable (for each observation) $\mathcal{Y}_{ji}$ contains $K_j$ possible outcomes.\footnote{These categorical variables may have different numbers of outcomes, hence the indexing by $j$.} 
\color{black}	

Let $\mathrm{D_{ijk}}$ be an indicator variable equal to unity, if respondent $i$ gives the $k$'th response to
the $j'$th variable, and equals zero otherwise:
\begin{IEEEeqnarray}{lCr}
\mathrm{D_{ijk}}=\begin{cases}
1 & \text{if } \mathcal{Y}_{ji}=k \\
0 & \text{otherwise}
\end{cases}
\end{IEEEeqnarray}	

Next, we construct the recruitment (response) probabilities; each denotes the probability of observing a specific response pattern, given the class membership.

\subsubsection{The recruitment (response) probabilities}\label{Recruitment:Classes}
Let $\pi_{jtk}$ be the probability that an observation in class $t$ produces the $k$'th outcome on the $j$'th variable.	The recruitment probabilities are class-dependent, but are assumed to be homogeneous within classes, which implies that the following must hold: 
\begin{IEEEeqnarray}{lCr}
\pi_{jtk}\equiv\mathrm{Pr}(\mathcal{Y}_j=k|\Psi_{\mathlcal{k}}=t,  \B{\mathrm{x}}=\B{x}_g)\\  \hspace{2em}=\mathrm{Pr}(\mathcal{Y}_j=k|\Psi_{\mathlcal{k}}=t).\nonumber	
\end{IEEEeqnarray}	    
Under conditional independence, which is a necessary condition for the class membership identification, the manifest variables are independent of each other, given the class membership and the group's observables characteristics. The probability that observation $i$ in class $t$ produces a particular set of $J$ outcomes on the observed categorical variables is the product:
\begin{IEEEeqnarray}{lCr}\label{cond:prob:latent}
\mathrm{Pr}(\mathcal{Y}_1=\mathlcal{y}_{1i},...,\mathcal{Y}_J=\mathlcal{y}_{Ji}|\Psi_{\mathlcal{k}}=t, \B{\mathrm{x}}=\B{x}_g)\\=\prod_{j=1}^{J}\mathrm{Pr}(\mathcal{Y}_j=\mathlcal{y}_{ji}|\Psi_{\mathlcal{k}}=t, \B{\mathrm{x}}=\B{x}_g)\nonumber\\
=\prod_{j=1}^{J}\mathrm{Pr}(\mathcal{Y}_j=\mathlcal{y}_{ji}|\Psi_{\mathlcal{k}}=t)=\prod_{j=1}^{J}\prod_{k=1}^{K_j}(\pi_{jtk})^{D_{ijk}}\nonumber
\end{IEEEeqnarray}

For any given class $t$ and observed categorical variable $j$, the following requirement must be satisfied $\sum_{k=1}^{K_j}\pi_{jtk}=1$.

The probability density function across all classes is the total probability over the conditional probability in \eqref{cond:prob:latent}:
\begin{IEEEeqnarray}{lCr}\label{Total:Recruitment:Prob}
\mathrm{Pr}(\mathcal{Y}_1=\mathlcal{y}_{1i},...,\mathcal{Y}_J=\mathlcal{y}_{Ji}|\B{\mathrm{x}}=\B{x}_g)\\=\sum_{t=1}^{\mathlcal{k}}\mathrm{Pr}(\mathcal{Y}_1=\mathlcal{y}_{1i},...,\mathcal{Y}_J=\mathlcal{y}_{Ji}, \Psi_{\mathlcal{k}}=t|\B{\mathrm{x}}=\B{x}_g)\nonumber\\
=\sum_{t=1}^{\mathlcal{k}} \mathrm{Pr}(\Psi_{\mathlcal{k}}=t|\B{\mathrm{x}}=\B{x}_g)\nonumber\\ \hspace{2em}\times \mathrm{Pr}(\mathcal{Y}_1=\mathlcal{y}_{1i},...,\mathcal{Y}_J=\mathlcal{y}_{Ji}| \Psi_{\mathlcal{k}}=t,\B{\mathrm{x}}=\B{x}_g)\nonumber\\
=\sum_{t=1}^{\tilde{G}}\lambda_{t|\B{x}_g}\prod_{j=1}^{J}\prod_{k=1}^{K_j}(\pi_{jtk})^{D_{ijk}}\nonumber
\end{IEEEeqnarray}
where the parameters to be estimated by the latent class model are $\lambda_{t|\B{x}_g}$ and $\pi_{jtk}$.

\subsubsection{A posteriori distribution function of The latent labels}\label{Posterior:Classes}
Given the prior and the recruitment probabilities' estimates for $\widehat{\lambda}_t$ and  $\widehat{\pi}_{jtk}$, respectively, the posterior probability that a given individual belongs to a given class, conditional on the observed response pattern $[\mathlcal{y}_{1i},...,\mathlcal{y}_{Ji}]$ is: 
\begin{IEEEeqnarray}{lCr}\label{posterior:latent:class}
\widehat{\mathrm{Pr}}(\Psi_{\mathlcal{k}}=t|\mathlcal{y}_{1i},...,\mathlcal{y}_{Ji},\B{x}_g)\\=\frac{\widehat{\lambda}_{t|\B{x}_g} \widehat{\mathrm{Pr}}(\mathcal{Y}_1=\mathlcal{y}_{1i},...,\mathcal{Y}_J=\mathlcal{y}_{Ji}|\Psi_{\mathlcal{k}}=t)}{\sum_{q=1}^{\tilde{G}}\widehat{\lambda}_{q|\B{x}_g} \widehat{\mathrm{Pr}}(\mathcal{Y}_1=\mathlcal{y}_{1i},...,\mathcal{Y}_J=\mathlcal{y}_{Ji}|\Psi=q)}\nonumber
\end{IEEEeqnarray}
where $t\in\curly{1,...,\mathlcal{k}}$.

The log-likelihood function to be maximized, with respect to the parameters values in the prior and the  recruitment probabilities $\lambda_{t|\B{x}_g}$ and $\pi_{jtk}$ using \eqref{Total:Recruitment:Prob} is:
\begin{IEEEeqnarray}{lCr}\label{loglikelihood:latent:class}
\ln L=\sum_{i=1}^{N}\ln\braket{\sum_{t=1}^{\tilde{G}}\lambda_{t|\B{x}_g}\prod_{j=1}^{J}\prod_{k=1}^{K_j}(\pi_{jtk})^{D_{ijk}}}
\end{IEEEeqnarray}
where the estimation procedure is expectation-maximization (EM) algorithm \cite{dempster1977maximum}.\footnote{The EM algorithm enables us to maximize the log-likelihood function in \eqref{loglikelihood:latent:class}, iteratively, to simplify the estimation process. Moreover, in the absence of slope covariates in both the class-prior and class-conditional probability functions, these probabilities are estimated nonparametrically. However, in the present case, the class-prior probability functions are estimated parametrically, due to the non-random assignment embedded in the presence of covariates.
	This is important for satisfying the non-covariate shift notion, as has been discussed earlier. In an important paper by \cite{greene2003latent}, a similar likelihood function is maximized, using a parametric technique.}

This log-likelihood function is identical in form to the standard finite mixture model log-likelihood. As with any finite mixture model, the EM algorithm
is applicable, because each individual's class membership is unknown and may be treated as
missing data \cite{mclachlan2000mixtures,mclachlan2007algorithm}.

The EM algorithm is an iterative procedure involving two sequential steps: an expectation and maximization. First, initial parameter values $\widehat{\lambda}_t^{\mathrm{old}}$ are arbitrarily chosen and $\widehat{\pi}_{jtk}^{\mathrm{old}}$ for each $t\in\curly{1,..,\mathlcal{k}}$ and $j\in\curly{1,..,J}$ for all $k\in\curly{1,..,K_j}$.
In the expectation step, calculate the "missing" class membership probabilities using  \eqref{posterior:latent:class}:
\newcommand\Resize[1]{\scaleobj{0.8}{\B{\mathlcal{{#1}}}}}
\begin{IEEEeqnarray}{lCr}\label{EM:Expectation}
\\ \scaleobj{1.4}{\mathfrak{P}}_t(\Resize{X}_g,\Resize{Y}_i) \equiv\widehat{\mathrm{Pr}}(\Psi_{\mathlcal{k}}=t|\mathlcal{y}_{1i},...,\mathlcal{y}_{Ji},\B{x}_g)\nonumber\\=\frac{\widehat{\lambda}_{t|\B{x}_g}^{\mathrm{old}} \prod_{j=1}^{J}\prod_{k=1}^{K_j}(\widehat{\pi}_{jtk}^{\mathrm{old}})^{D_{ijk}} }{\sum_{q=1}^{T}\widehat{\lambda}_{q|\B{x}_g}^{\mathrm{old}} \prod_{j=1}^{J}\prod_{k=1}^{K_j}(\widehat{\pi}_{jqk}^{\mathrm{old}})^{D_{ijk}}}\nonumber
\end{IEEEeqnarray}

In the maximization step, we update
the parameter estimates by maximizing the log-likelihood function in \eqref{loglikelihood:latent:class}, given the estimated posterior in \eqref{EM:Expectation}. The new-prior probabilities are:
\vspace{0.5em}
\begin{IEEEeqnarray}{lCr}\label{EM:Maximization:first}
 \hat\lambda_{t|\B{x}_g}^{\mathrm{new}}=\frac{1}{N}\sum_{i=1}^{N}\widehat{\mathrm{Pr}}(\Psi_{\mathlcal{k}}=t|\mathlcal{y}_{1i},...,\mathlcal{y}_{Ji},\B{x}_g) 
\end{IEEEeqnarray}
and the new class conditional probabilities are:		
\begin{IEEEeqnarray}{lCr}\label{EM:Maximization:last}
 \hat\pi_{jtk}^{\mathrm{new}}=\frac{\sum_{i=1}^{N}D_{ijk}\widehat{\mathrm{Pr}}(\Psi_{\mathlcal{k}}=t|\mathlcal{y}_{1i},...,\mathlcal{y}_{Ji},\B{x}_g)}{\sum_{i=1}^{N}\widehat{\mathrm{Pr}}(\Psi_{\mathlcal{k}}=t|\mathlcal{y}_{1i},...,\mathlcal{y}_{Ji},\B{x}_g)}.
\end{IEEEeqnarray}

We replace the old estimates $\widehat{\lambda}_t^{\mathrm{old}}$ and $\widehat{\pi}_{jtk}^{\mathrm{old}}$ with the new estimates $\widehat{\lambda}_t^{\mathrm{new}}$ and $\widehat{\pi}_{jtk}^{\mathrm{new}}$, respectively, and repeat the expectation and maximization steps in \eqref{EM:Expectation}-\eqref{EM:Maximization:last}, until a convergence criterion is satisfied for these new parameter values.

Using the estimated posterior function in \eqref{EM:Expectation}, the sequence of fitted labels $\widehat{\psi}_{i,g}$ are the arguments that maximize ($\arg \max$) the estimated posterior distribution function. Thus, given a response pattern $\Resize{Y}_i=[\mathlcal{y}_{1i},...,\mathlcal{y}_{Ji}]$ and group's observed characteristics $\B{x}_g$, the fitted label for the latent $i$'th datum is:
\begin{IEEEeqnarray}{lCr}\label{fitted:labels}
\widehat{\psi}_{i,g}=\underset{t}{\arg\max}\hspace{1em}	\scaleobj{1.4}{\mathfrak{P}}_t(\Resize{X}_g,\Resize{Y}_i)
\end{IEEEeqnarray} 
	
Next, we utilize the experts' opinions  in each reference group to evaluate the participants' share. The reference groups are identified by using both the group's observed characteristics and the fitted labels in \eqref{fitted:labels}, capturing its unobserved characteristics.

\subsection{The opinion space}\label{Sec:Opinions:Experts}
We introduce an opinion space composed of a set of experts defined as:
\begin{IEEEeqnarray}{rCl}
\scaleobj{0.9}{\Lambda=\curly{(\SizeSym{X},\SizeSym{Y},\psi^E)\hspace{0.5em}|\hspace{0.5em}\SizeSym{X}\in\mathbb{R}^{L_{\B{x}}},\hspace{0.5em}\SizeSym{Y}\in \mathfrak{Y}, \hspace{0.5em}\psi^E=1,...,\mathlcal{k}}},\nonumber\end{IEEEeqnarray} in which an expert $\varphi\in\Lambda$, a set of observed characteristics $(\SizeSym{X},\SizeSym{Y})$ and unobserved characteristics ($\psi^E$),  has a discretized opinion $\chi_{\varphi}\in\curly{0,1}$ regarding the expected participation decision of a member belonging to own reference group.\footnote{An expert opinion reflects the decision that a member of his group is more likely to make. That is, being a participant or a non-participant.}

Let $\varphi_e=(\varphi_e^*,\psi_e^E)\in\Lambda$, where $\varphi_e^*=(\SizeSym{X}_e,\SizeSym{Y}_e)$. Denote a random sample $\Lambda^S=\curly{\varphi_e^*,\chi_{\varphi_e}}_{e=1}^{N^E}$ consisting of $N^E$ experts $\curly{\varphi_e^*}_{e=1}^{N^E}$  and their respective opinions $\curly{\chi_{\varphi_e}}_{e=1}^{N^E}$, where  $\chi_{\varphi_e}\in\curly{0,1}$. Each of the opinions in $\curly{\chi_{\varphi_e}}_{e=1}^{N^E}$ is an independent realization of a Bernoulli random variable $\mathlarger{\Bigger{\omega}}$, with probability of success defined by the function $m^{\mathlcal{e}}(\SizeSym{X}_e, \psi_e^E)$.\footnote{Not to be confused with the actual participants' share $m(\B{x}_g^E,\psi_{i,g}^E)$  depicted in \eqref{Equilibrium:Condition}.} 

It follows that $\Lambda^S$ consists entirely of the experts' observed characteristics and their opinions. The unobserved characteristics are essential for being able to assign the experts into their respective reference groups. However, the unobserved and observed characteristics are interrelated, through the a posteriori probability density function depicted in \eqref{EM:Expectation}. The former are substituted with their fitted values, which are the arguments maximizing the posterior probability density function, given the observed characteristics.
Using the aforementioned interrelationship and given the sample $\Lambda^S$,  the set of expert opinions that are assigned to latent class $t$ is denoted by: 
\begin{IEEEeqnarray}{lCr}
\mathcal{O}_{p_{(t)}}=\LeftCurly{1.8}{\chi_{\varphi_e}\hspace{0.5em}\Bigger{|}\hspace{0.5em}(\SizeSym{X}_e,\SizeSym{Y}_e,\chi_{\varphi_e})\in\Lambda^S},\\ \hspace{0.5em}\RightCurly{1.8}{t\in\underset{\psi}{\arg\max}\hspace{0.5em}\scaleobj{1.4}{\mathfrak{P}}_{\psi}(\SizeSym{X}_e,\SizeSym{Y}_e)}	
\nonumber
\end{IEEEeqnarray}

 The entire experts' opinions data set is denoted by the sequence $\curly{\mathcal{O}_{p_{(t)}}}_{t=1}^{\mathlcal{k}}$.
\newcommand\RandX{\scaleobj{1.2}{\B{\mathrm{x}}}}

The participants' shares given $\mathlcal{k}$ latent classes are obtained by Bayes' rule:\footnote{The expression $\mathrm{Pr}\close{\Psi_{\mathlcal{k}}=t}$ is canceled out and thus, is not presented in either the numerator or the denominator in \eqref{Prob:Expert}. }
\begin{IEEEeqnarray}{lCr}\label{Prob:Expert}
m_{\mathlcal{k}}^e(\SizeSym{X},t)\\=\scaleobj{0.8}{\fracBig{f_{\RandX|\Bigger{\omega}=1,\Psi_{\mathlcal{k}}=t}\close{\SizeSym{X}|\Bigger{\omega}=1,\Psi_{\mathlcal{k}}=t}\mathrm{Pr}\close{\Bigger{\omega}=1|\Psi_{\mathlcal{k}}=t}}{\sum_{\omega=0}^{1}f_{\RandX|\Bigger{\omega}=\omega,\Psi_{\mathlcal{k}}=t}\close{\SizeSym{X}|\Bigger{\omega}=1,\Psi_{\mathlcal{k}}=t}\mathrm{Pr}\close{\Bigger{\omega}=\omega|\Psi_{\mathlcal{k}}=t}}}\nonumber
\end{IEEEeqnarray}
where $m_{\mathlcal{k}}^e(\SizeSym{X},t)\equiv
\mathrm{Pr}\close{\Bigger{\omega}=1|\RandX=\SizeSym{X},\Psi_{\mathlcal{k}}=t}$.

Neither of the density functions $f_{\RandX|\Bigger{\omega}=0,\Psi_{\mathlcal{k}}=t}$ nor $f_{\RandX|\Bigger{\omega}=1,\Psi_{\mathlcal{k}}=t}$ is known or specified by the researcher, and they are substituted with their respective estimates: $\widehat{f}_{\B{\mathrm{x}}|\Bigger{\omega}=0}^{t}$ and $\widehat{f}_{\B{\mathrm{x}}|\Bigger{\omega}=1}^{t}$, as described in \eqref{Density:X:estimate}, to follow.  Similarly, the probabilities $\mathrm{Pr}\close{\Bigger{\omega}=1|\Psi_{\mathlcal{k}}=t}$ and $\mathrm{Pr}\close{\Bigger{\omega}=0|\Psi_{\mathlcal{k}}=t}$ are replaced by their estimates $\mathfrak{p}_t$ and $1-\mathfrak{p}_t$, respectively. Thus,
\begin{IEEEeqnarray}{lCr}\label{Smoothing:Average:Opinion}
\scaleobj{0.8}{\widehat{m}_{\mathlcal{k}}^e(\SizeSym{X},t)=\fracBig{\mathfrak{p}_t\widehat{f}_{\B{\mathrm{x}}|\Bigger{\omega}=1}^{t}(\SizeSym{X})}{\mathfrak{p}_t\widehat{f}_{\B{\mathrm{x}}|\Bigger{\omega}=1}^{t}(\SizeSym{X})+(1-\mathfrak{p}_t)\widehat{f}_{\B{\mathrm{x}}|\Bigger{\omega}=0}^{t}(\SizeSym{X})}},\hspace{3.5em}\\ \hspace{0.5em} \scaleobj{0.8}{\mathfrak{p}_t=\frac{1}{N_t^E}\sum_{\chi_{\varphi_e}\in\mathcal{O}_{p_{(t)}}}\chi_{\varphi_e}}\nonumber
\end{IEEEeqnarray}
where $N_{t}^E$ is the cardinality (number of elements) of the set $\mathcal{O}_{p_{(t)}}$.

Using the Parzen-Rosenblatt \cite{rosenblatt1956remarks,parzen1962estimation} window method for a nonparametric density estimation given a $L_{\B{x}}\times L_{\B{x}}$  bandwidth  matrix $\mathcal{H}$,\footnote{The multivariate gaussian kernel density estimator is employed due to its applicability to multivariate data. Unlike in the case of semiparametric estimation, in the case of nonparametric estimation there is a ``protocol'' for finding the optimal bandwidth for instance,   \cite{scott1991feasibility}'s rule. } we denote a conditional density estimator of  the random variable vector $\B{\mathrm{x}}\in \mathbb{R}^{L_{\B{x}}}$ given the opinion $\omega\in\curly{0,1}$ and an estimated membership in latent class $t$:
\begin{IEEEeqnarray}{lCr}\label{Density:X:estimate}
\widehat{f}_{\B{\mathrm{x}}|\Bigger{\omega}=\omega}^{t}(\B{x})=\frac{1}{N_{t,\omega}^E(2\pi)^{\frac{L_{\B{x}}}{2}}}\abs{\mathcal{H}}^{-1/2}\\ \times \sum_{\SizeSym{X}\in\mathcal{O}_{\B{x}_{(t)}}^{\omega}}\exp\curly{\frac{1}{2}\close{\SizeSym{X}-\B{x}}^T\mathcal{H}^{-1}\close{\SizeSym{X}-\B{x}}}\nonumber
\end{IEEEeqnarray}
where \begin{IEEEeqnarray}{lCr} \mathcal{O}_{\B{x}_{(t)}}^{\omega}=\LeftCurly{1.8}{\SizeSym{X}_e\hspace{0.5em}\Bigger{|}\hspace{0.5em}(\SizeSym{X}_e,\SizeSym{Y}_e,\chi_{\varphi_e})\in\Lambda^S},\\ \hspace{0.5em}\RightCurly{1.8}{t\in\underset{\psi}{\arg\max}\hspace{0.5em}\scaleobj{1.4}{\mathfrak{P}}_{\psi}(\SizeSym{X}_e,\SizeSym{Y}_e),\hspace{0.5em}\chi_{\varphi_e}=\omega}\nonumber
\end{IEEEeqnarray} is a subset of $\curly{\SizeSym{X}_e}_{e=1}^{N^E}$, consisting only of the observed characteristics related to experts assigned to latent class $t$ with the opinion $\omega$ and $N_{t,\omega}^E$ is the cardinality (number of elements) of the set $\mathcal{O}_{\B{x}_{(t)}}^{\omega}$. The determinant of $\mathcal{H}$ is $\abs{\mathcal{H}}$.
The main idea behind the mapping from an expert set to a sequence of opinions is to take advantage of auxiliary data (e.g., survey data, training data and the like), in which each data point depicts an opinion of a specific expert. Averaging the opinions in each reference group obtained from \eqref{Smoothing:Average:Opinion} generates the share of participants belonging to that reference group. Thus, the best forecast, resulting from the various reference groups is a refinement of the wisdom of crowd (\cite{galton1907vox,surowiecki2005wisdom,prelec2017solution,hummel1996statistical,budescu2014identifying}). The type and number of reference groups are unobserved and are estimated by the posterior class membership probability density function.

The proposed implementation relies on the utilization of two data sets: (i) a survey data set consisting of experts' opinions $\curly{\mathcal{O}_{p_{(t)}}}_{t=1}^{\mathlcal{k}}$, with the observed group-level covariates; and (ii) a truncated data set consisting of both individual-level covariates as well as group-level covariates. The proposed procedure is closely related to the similarity-based classification, which is referred to as ``nearest neighbor algorithm'', in the field of machine learning (e.g., \cite{chen2009similarity}). Nearest neighbor algorithm assigns labels in the truncated (test) data set based on the similarities between this data set and the non-truncated labeled (training) data set. However, in the present case, the labels are unobserved not only in the truncated data set, but in both data sets.\footnote{This phenomenon is termed ``unlabeled data'' in the field of machine learning.} Thus, the purpose of the survey data set is to estimate the posterior distribution function in order to fit the labels in the truncated data set.  

\subsection{Estimating the substantive equation}\label{Partial:Index:Model:Estimation}
We formulate the estimation procedure in terms of a non-linear least squares (NLS) minimization. Although the substantive equation is a linear function of its covariates, it can be reformulated as a partially linear single-index model in order to correct for the endogenous selection bias. The single-index modeling draws on the \cite{johnson1984extensions} Lemma, alleviating the complexity present in high-dimension covariates space. In our model, the single index function is referred to as the bias term \cite{heckman1979sample} and is constructed using \eqref{Latent:substantive:equation}, by taking its conditional expectation, given the covariates and being a participant:
\begin{IEEEeqnarray}{lCr}\label{Expected:Latent:substantive:equation}
\mathbb{E}[y_{1i,g}^{*}|\mathrm{S}=1,\B{w}_i,\B{z}_i,\B{x}_g,\psi_{i,g}]\\=\B{w}_i^T\B{\theta}+\mathbb{E}[\disturb_{1i}|\mathrm{S}=1,\B{w}_i,\B{z}_i,\B{x}_g,\psi_{i,g}]\nonumber
\end{IEEEeqnarray}
where $\B{w}_i$ is a vector of the substantive equation's covariates; while $\B{x}_i$ and $\B{z}_i$ are specific group-level and individual-level characteristics, respectively.

The residual $\epsilon_i$ between $y_{1i,g}$ and its conditional expectation, given participation \eqref{Expected:Latent:substantive:equation} in the truncated data is constructed as:
\begin{IEEEeqnarray}{lCr}
\epsilon_i={y_{1i,g}}-\mathbb{E}[y_{1i,g}^{*}|\mathrm{S}=1,\B{w}_i,\B{z}_i,\B{x}_g,\psi_{i,g}]
\end{IEEEeqnarray}
Using \eqref{Expected:Latent:substantive:equation} and denoting $\mathbb{E}[\disturb_{1i}|\mathrm{S}=1,\B{w}_i,\B{z}_i,\B{x}_g,\psi_{i,g}]\equiv\mathcal{M}(\beta  m^{\mathlcal{e}}(\B{x}_g,\psi_{i,g})+\B{z}_i^T\B{\eta}+\close{\B{x}_{g}^{c}}^T\B{\delta})$ we arrive at the partially linear single-index model:
\begin{IEEEeqnarray}{lCr}\label{index:regression}
y_{1i,g}=\B{w}_i^T\B{\theta}+\mathcal{M}\LeftClose{1.2}{\beta m^{\mathlcal{e}}(\B{x}_g,\psi_{i,g})+\B{z}_i^T\B{\eta}}\\+\RightClose{1.2}{\close{\B{x}_{g}^{c}}^T\B{\delta})+\epsilon_i} \nonumber
\end{IEEEeqnarray}
where  $\B{x}_{g}^{c}$ are the contextual covariates.

\vspace{0.5em}
However, neither the function $m^{\mathlcal{e}}(\cdot)$ nor its $\psi_{i,g}$ argument is observed. Thus, they are substituted with their respective estimates  $\widehat{m}_{\mathlcal{k}}^e(\cdot)$ and $\widehat{\psi}_{i,g}$ given $\mathlcal{k}$ possible latent classes (labels), obtained from the survey data or any other auxiliary data. The former is constructed using \eqref{Smoothing:Average:Opinion}, which is  a refinement of the vox populi (average forecast) mechanism (see section \ref{Sec:Opinions:Experts}):
\begin{IEEEeqnarray}{lCr}
\scaleobj{0.8}{\widehat{m}_{\mathlcal{k}}^e(\B{x}_g,\widehat{\psi}_{i,g})=\fracBig{\mathfrak{p}_{\widehat{\psi}_{i,g}}\widehat{f}_{\B{\mathrm{x}}|\Bigger{\omega}=1}^{\widehat{\psi}_{i,g}}(\B{x}_g)}{\mathfrak{p}_{\widehat{\psi}_{i,g}}\widehat{f}_{\B{\mathrm{x}}|\Bigger{\omega}=1}^{\widehat{\psi}_{i,g}}(\B{x}_g)+(1-\mathfrak{p}_{\widehat{\psi}_{i,g}})\widehat{f}_{\B{\mathrm{x}}|\Bigger{\omega}=0}^{\widehat{\psi}_{i,g}}(\B{x}_g)}},\nonumber\\ \hspace{0.5em}\scaleobj{0.8}{ \mathfrak{p}_{\widehat{\psi}_{i,g}}=\frac{1}{N_{{\widehat{\psi}_{i,g}}}^E}\hspace{1em}\sum_{\chi_{\varphi_e}\in\mathcal{O}_{p_{({\widehat{\psi}_{i,g}})}}}\chi_{\varphi_e}}\nonumber
\end{IEEEeqnarray}
where $N_{\widehat{\psi}_{i,g}}^E$ is the cardinality (number of elements) of the set $\mathcal{O}_{p_{(\widehat{\psi}_{i,g})}}$ and $\widehat{\psi}_{i,g}=\underset{t}{\arg\max}\hspace{0.5em}\mathfrak{P}_t(\B{x}_g,\scaleobj{0.8}{\mathlcal{Y}}_i)$.

Our objective is to estimate the substantive equation \eqref{index:regression}, which includes the function $\mathcal{M}(.)$ as an additional covariate controlling for the endogenous sample selection. However, the function  $\mathcal{M}(.)$  in  \eqref{index:regression} is unknown and has to be approximated. 
In the next section we attend to this issue.

\subsection{Transformation of both Cosine and Fourier series for unknown \\\text{ } \hspace{1.6em} functions estimation}\label{Sec:Esitmate:Fourier}

The function $\mathcal{M}(.)$ in \eqref{index:regression} is approximated using its conditional moment expansion by employing either The Cosine or The Fourier sequence.  The Cosine sequence requires that the support of the index variable in \eqref{index:regression} will be on the $[0,1]$ domain, while the Fourier sequence requires that the support will be on the $[-1,1]$ domain.\footnote{Fourier series decomposes a ``periodic'' signal into a sum of an infinite number of harmonics (sine and cosine functions) of different frequencies and  amplitudes, while Fourier transform decomposes a ``non-periodic'' signal into an infinite number of harmonics having different frequencies and amplitudes. } This assumption does not entail loss of generality, because it is satisfied by utilizing a different monotone transformation function on the index variable \cite{horowitz2014adaptive} in each one of the Cosine and Fourier series. The series generated by the transformation is referred to as a transformed Cosine (or Fourier) series. 

In the case of the (transformed) Cosine sequence the conditional moment expansion of $\mathcal{M}(.)$ is denoted by $\widehat{\mathcal{M}}^c(\mathlcal{b};\B{\vartheta_{c}})$ and is defined $\forall\mathlcal{b}\in\mathbb{R}$ as:
\begin{IEEEeqnarray}{lCr}
\widehat{\mathcal{M}}^c(\mathlcal{b};\B{\vartheta_{c}})=\alpha_{c}+\sum_{k=1}^{\mathcal{K}}\tau_{k}^c\cos(\varphi(\mathlcal{b})\pi k)
\end{IEEEeqnarray}
where $\varphi(.)$ is some known, arbitrarily chosen, strictly monotonic twice differentiable mapping $\mathbb{R}\mapsto(0,1)$, $\B{\vartheta_{c}}\equiv\curly{\alpha_{c},\B{\tau^c}}$ and $\B{\tau^c}\equiv[\tau_{1}^c,...,\tau_{\mathcal{K}}^c]$, with $\mathcal{K}$ being the number of elements in the expansion.

Similarly, in the case of the (transformed) Fourier sequence the conditional moment expansion of $\mathcal{M}(.)$ is denoted by $\widehat{\mathcal{M}}^f(\mathlcal{b};\B{\vartheta_{c}})$ and is defined $\forall\mathlcal{b}\in\mathbb{R}$ as:
\begin{IEEEeqnarray}{lCr}
\widehat{\mathcal{M}}^f(\mathlcal{b};\B{\vartheta_{f}})=\alpha_{f}+\sum_{k_1=1}^{\mathcal{K}}\tau_{1_{k_1}}^f\cos(\zeta(\mathlcal{b})\pi k_1)\\+\sum_{k_2=1}^{\mathcal{\mathcal{K}}}\tau_{2_{k_2}}^f\sin(\zeta(\mathlcal{b})\pi\nonumber k_2)
\end{IEEEeqnarray}
where $\zeta(.)$ is some known, arbitrarily chosen, strictly monotonic twice differentiable mapping $\mathbb{R}\mapsto(-1,1)$,\footnote{The main drawback of Fourier series, however, is the requirement of the approximated
	function to be periodic on a bounded interval. This is problematic, as we are interested in approximating a non-periodic function defined on an unbounded interval. To alleviate this problem, we use monotonic mapping of the function's argument from the real line to the [-1,1] domain to make it periodic only at infinity and bounded on this domain. The aforementioned transformation results in enhanced accuracy of the estimates. due to the flexibility of Fourier series estimator, without being restricted to the family of periodic functions.} $\B{\vartheta_{f}}\equiv\curly{\alpha_{f},\B{\tau_{1}^f},\B{\tau_{2}^f}}$ and $\B{\tau_{m}^f}\equiv[\tau_{m_1}^f,...,\tau_{m_{\mathcal{K}}}^f]$, $m=1,2$ representing Sine or Cosine, respectively.

For brevity, we denote the parameter vector $\B{\theta}^*\equiv\braket{\B{\theta}^T,\B{\eta}^T,\B{\delta}^T,\beta}^T$.
Following \cite{racine2014oxford}, given the non-linear function $\widehat{\mathcal{M}}^{\mathcal{G}}(\mathlcal{b};\B{\vartheta_{\mathcal{G}}})$ with $\mathcal{G}\in\curly{c,f}$ an index model can be estimated as follows: 
\begin{IEEEeqnarray}{lCr}\label{Objective:Norm2:function}
\\(\B{\widehat{\theta^*},\widehat{\vartheta_{\mathcal{G}}}})=\arg\underset{\B{(\theta^*,\vartheta_{\mathcal{G}})}\in\Theta\times \Delta_{\mathcal{K}}}{\min}\frac{1}{n}\sum_{i=1}^{n}\LeftClose{1.8}{y_{1i,g}-\B{w}_i^T\B{\theta}}\nonumber\\-\RightClose{1.8}{\widehat{\mathcal{M}}^{\mathcal{G}}(\beta  \widehat{m}_{\mathlcal{k}}^e(\B{x}_g,\psi_{i,g})+\B{z}_i^T\B{\eta}+\close{\B{x}_{g}^{c}}^T\B{\delta};\B{\vartheta_{\mathcal{G}}})}^2\nonumber
\end{IEEEeqnarray}
where $y_{1i,g}$ is the substantive equation's dependent variable; $\mathcal{K}$ is the number of elements in the expansion; $\B{w_i}$ and $\B{\theta}$ stand for the covariates set and the parameter set, respectively in the linear part of the substantive equation. Note that the combination in \eqref{Objective:Norm2:function} of the linear component $\B{w}_i^T\B{\theta}$ and the non-linear component $\mathcal{M}(\cdot)$ implies partial linearity of the model.\footnote{This is where we depart from \cite{racine2014oxford}, who introduce only the non-linear component, as they did not deal with truncation.}

We require that the expectation of the objective function in \eqref{Objective:Norm2:function} is finite for all values of the parameters $\B{(\theta^*,\vartheta_{\mathcal{G}})}$\footnote{This assumption can be relaxed using a positive weight function $\mathcal{K}(x)$ on $(0,\infty)$ in the nonlinear minimization (see, \cite{racine2014oxford}).} that is,
\begin{IEEEeqnarray}{lCr}
\mathbb{E}\LeftBraket{2.5}{\underset{\B{(\theta^*,\vartheta_{\mathcal{G}})}\in\Theta\times \Delta_{\mathcal{K}}}{\sup}\frac{1}{n}\sum_{i=1}^{n}\LeftClose{1.8}{y_{1i,g}-\B{w}_i^T\B{\theta}}}\\-\RightBraket{2.5}{\RightClose{1.8}{\widehat{\mathcal{M}}^{\mathcal{G}}(\beta  \widehat{m}_{\mathlcal{k}}^e(\B{x}_g,\psi_{i,g})+\B{z}_i^T\B{\eta}+\close{\B{x}_{g}^{c}}^T\B{\delta};\B{\vartheta_{\mathcal{G}}})}^2} < \infty.\nonumber
\end{IEEEeqnarray}

Next we have to modify \eqref{Objective:Norm2:function} and accommodate it for the presence of latent reference groups. This is done by introducing a penalization into the model.

\section{The optimal number of latent reference groups}\label{Unknown:Latent:Groups}
In practice, the number of latent classes (labels) in the truncated data is unknown. Arbitrarily choosing the number of latent classes may amount to misspecification.\footnote{A non-feasible solution is to assume that any individual observation is its own advisor (reference group) based on his past experience. This is problematic (unless an auxiliary data set with historical individual level participation probabilities is accessible), as the individual data consists of participants only, and consequently one cannot estimate the probability to participate for a specific data-point using only one observation, which is the participant herself.} To alleviate probable misspecification, we propose an estimation procedure generating the ``optimal'' number of latent classes to fit the correct model without arbitrarily assuming the number of reference groups. This procedure specifies the participation decision that best fits the data generation process in the truncated data, which is related to some criterion function (to be defined in \eqref{SPLM:Criteria} to follow). The aforementioned participation decision is chosen from a menu consisting of selection equations differentiated by $\mathlcal{k}$ number of available (latent) reference groups. This is achieved by minimizing an additive penalized objective function $\Upsilon(\B{\varphi})$ \cite{racine2014oxford}:
\begin{IEEEeqnarray}{lCr}\label{Penalization}
\Upsilon(\B{\varphi})=\text{Loss function}(\B{\varphi}) + \text{Penalty function}(\B{\varphi},\lambda_n)\hspace{1em}
\end{IEEEeqnarray} 
where $\B{\varphi}$ is a vector of estimated parameters and $\lambda_n$ is a tuning parameter.\footnote{When $\lambda_n$ approaches zero the penalty function is not effective, leading to the parameter estimates that would have been obtained without penalization.}

We note that increasing the number of reference groups decreases the model bias, due to enhanced information (explanatory ability), however, it is at a cost of higher variance in the model (low accuracy). To overcome this bias-variance trade-off, a penalization procedure is applied, as is depicted by the penalty function in \eqref{Penalization}. The penalized regression is also termed ``sparse regression'', where ``sparsity'' implies that only a small fraction of the predictor variables has an influence on the dependent variable \cite{friedman2012fast}. These regression methods are intended to find the subset of the most influent predictors by shrinking down the parameter estimates toward zero and reducing the number of non-zero parameter estimates. 

The most popular choice of loss-functions  are Mean Squared Error (MSE), negative log-likelihood and profiled least squares. In our case, we employ the Mean Squared Error (MSE) loss function to be consistent with the nonlinear least squares problem depicted in \eqref{Objective:Norm2:function}.
In order to select the optimal number of latent reference groups, we use the SCAD penalty function, as it  nests the LASSO as a special case, defined as:\footnote{
	The SCAD performs well in partially linear index models \cite{racine2014oxford,liang2010estimation}.}
\begin{IEEEeqnarray}{lCr}
p_{\lambda_n}(v)=\begin{cases}
\lambda_n v & \text{if } 0\le v\le \lambda_n \\
-\frac{v^2-2a\lambda_n v+\lambda_n^2}{2(a-1)} & \text{if } \lambda_n< v< a\lambda_n \\
\frac{(a+1)\lambda_n^2}{2} & \text{if } v> a\lambda_n \\
\end{cases}
\end{IEEEeqnarray}
where $a>2$ is a constant. For practical use we set $a=3.7$  \cite{racine2014oxford}.\footnote{This has been shown to facilitate computation time.}

Next, we present an algorithm for determining the optimal number of latent reference groups, employing the SCAD penalty function.

\subsection{Estimating the number of latent reference groups}
We construct a sequence of functions $\curly{\widehat{m}_{\mathlcal{k}}^e(\B{x}_g,\psi_{i,g})}_{\mathlcal{k}=1}^{\tilde{G}_{\max}}$, where $\tilde{G}_{\max}\in\mathbb{N}$ is the largest latent labels (outcomes) number, $\widehat{m}_{\mathlcal{k}}^e(\B{x}_g,\psi_{i,g})$ is the conditional participation probability, given $k\le \tilde{G}_{\max}$ latent labels, the observed group's characteristics $\B{x}_g$ and belonging to label (being a member in class) $\psi_{i,g}\in\curly{1,...,k}$.\footnote{In a recent contribution \cite{diebold2017beating} also utilize a penalty function to combine forecasts. However, they utilize the LASSO penalty function which is restrictive in that it forces most of the covariates to have zero coefficients, instead of allowing for a combination of the covariates to be utilized like the SCAD penalty function employed here. } For brevity, we define $\widehat{\rho}_{\mathlcal{k}_{i,g}}\equiv\widehat{m}_{\mathlcal{k}}^e(\B{x}_g,\widehat{\psi}_{i,g})$, where $\widehat{\psi}_{i,g}$ is the estimated label membership for the $i$'th observation, given $\mathlcal{k}$ latent classes. The partially linear single index regression is represented as:
\begin{IEEEeqnarray}{lCr}\label{Penalized:regression}
\scaleobj{0.8}{y_{_{1i,g}} = \mathcal{F}_{i,g}(\B{\varphi}) + \epsilon_i, \hspace{1em}} \\ \scaleobj{0.8}{ \mathcal{F}_{i,g}(\B{\varphi})=\B{w_i}^T\B{\theta}+\widehat{\mathcal{M}}^{\mathcal{G}}\close{\sum_{k=1}^{\tilde{G}}\beta_{k}\widehat{\rho}_{k_{i,g}}+\B{z}_i^T\B{\eta}+ \close{\B{x}_{g}^{c}}^T\B{\delta};\nonumber \hspace{0.5em}\B{\vartheta_{\mathcal{G}}}}}
\end{IEEEeqnarray}
where $\B{\varphi}=\curly{\B{\theta,\beta,\eta,\delta},\B{\vartheta_{\mathcal{G}}}}$, such that $\B{\beta}\equiv\curly{\beta_1,...,\beta_{\tilde{G}}}$. 

In the first step, a solution path $\B{\varphi}_{\lambda_n}=\LeftCurly{1.2}{\B{\theta_{\lambda_n},\beta_{\lambda_n},\eta_{\lambda_n}}}$, $\RightCurly{1.2}{\beta_{\delta_{\lambda_n}},\B{\vartheta_{\mathcal{G}}}_{\lambda_n}}$ indexed by a tuning parameter,  $\lambda_n$, is estimated as a penalized  partially linear single index model \cite{liang2010estimation}:\footnote{Unlike the (penalized) partially linear single index model (PLSIM) estimation procedure introduced by \cite{liang2010estimation} which utilizes kernel estimator (suffering from bandwidth selection consideration) to approximate the unknown function $\mathcal{M}$, we use a Sieve estimator.}
\begin{IEEEeqnarray}{lCr}\label{SPLM:SCAD}
\scaleobj{0.8}{\B{\varphi}_{\lambda_n}=\arg \min  \curly{\frac{1}{2}\underbrace{ \normsq{\B{y_{_1}}-\mathcal{F}(\B{\varphi}_{\lambda_n})}{2}}_{\text{sum of squares error term}}+\underbrace{n\sum_{k=1}^{\tilde{G}}p_{\lambda_n}(\abs{\beta_{k_{\lambda_n}}})}_{{\text{penalty term}}}}}
\end{IEEEeqnarray}
where $\B{\beta_{\lambda_n}}\equiv\braket{\beta_{1_{\lambda_n}},...,\beta_{J_{\lambda_n}}}^T$, $\B{y_{_1}}=[\B{y_{_{1,1}}^T},...,\B{y_{_{1,G}}^T}]^T$ such that $\B{y_{_{1,g}}}=[y_{_{11,g}},...,y_{_{1n_{g},g}}]^T$  and the $\norm{\B{\cdotp}}{2}$ is the usual $\ell_2$ (Euclidean) norm.\footnote{The $\ell_p$ norm definition is:
	\begin{IEEEeqnarray}{lCr}
	\norm{\B{b}}{p}=\close{\sum_{i=1}^{n}\abs{b_{i}}^p}^{1/p}.
	\end{IEEEeqnarray}} 

In the second step, a criterion $\mathcal{C}_p$ is computed for the solution path $\B{\widehat{\varphi}}_{\lambda_n}$. The conventionally chosen criterion is BIC (Bayesian information criterion \cite{greene2003latent}) computed as:
\begin{IEEEeqnarray}{lCr}\label{SPLM:Criteria}
\mathcal{C}_p(\lambda_n) = \log (\text{MSE}(\lambda_n)) + \frac{\log n}{n} df_{\lambda_n}
\end{IEEEeqnarray}
where $\text{MSE}(\lambda_n)=n^{-1}\sum_{i=1}^{n} \close{y_{1i,g}-\mathcal{F}_{i,g}(\B{\widehat{\varphi}}_{\lambda_n})}^2$ and $df_{\lambda_n}$ is the number of non-zero coefficients in $\B{\widehat{\varphi}}_{\lambda_n}$.

The algorithm for finding the correct model requires estimating  \eqref{SPLM:SCAD} repeatedly, each time given a different tuning parameter value $\lambda_n$, and computing $\text{MSE}(\lambda_n)$ in order to find $\lambda_n$ which minimizes \eqref{SPLM:Criteria}.

Estimating \eqref{SPLM:SCAD} involves the utilization of a non-convex penalty function optimization, which enhances computational complexity. To alleviate this complexity and without loss of accuracy, we transform the optimization problem into a constrained one with a convex penalty function \cite{figueiredo2007gradient}.

Thus, we introduce the parameter vectors $\B{\beta_{\lambda_n}^{+}}\equiv\braket{\beta_{1_{\lambda_n}}^{+},...,\beta_{J_{\lambda_n}}^{+}}^T$  and $\B{\beta_{\lambda_n}^{-}}\equiv\braket{\beta_{1_{\lambda_n}}^{-},...,\beta_{J_{\lambda_n}}^{-}}^T$ where $\beta_{k_{\lambda_n}}^{+}=\max\curly{0,\beta_{k_{\lambda_n}}}$ and  
$\beta_{k_{\lambda_n}}^{-}=\max\curly{0,-\beta_{k_{\lambda_n}}}$  $\forall k$
and make the following substitution:
\begin{IEEEeqnarray}{lCr}\label{Difference:Operator}
\B{\beta_{\lambda_n}}=\B{\beta_{\lambda_n}^{+}-\beta_{\lambda_n}^{-}},\hspace{1em}  \B{\beta_{\lambda_n}^{+}}\ge \B{0}, \hspace{1em} \B{\beta_{\lambda_n}^{-}}\ge \B{0} 
\end{IEEEeqnarray}

Using $\B{\beta_{\lambda_n}^{+}}$ and $\B{\beta_{\lambda_n}^{-}}$ the optimization becomes:
\begin{IEEEeqnarray}{lCr}\label{Problem:convex}
\\ \scaleobj{0.8}{\B{\varphi}_{\lambda_n^*}=\arg \min  \LeftCurly{2}{\frac{1}{2}\underbrace{ \normsq{\B{y_{_1}}-\mathcal{F}(\B{\varphi}_{\lambda_n^*})}{2}}_{\text{sum of squares error term}}}+\RightCurly{2}{\underbrace{n\sum_{k=1}^{J}p_{\lambda_n}(\beta_{k_{\lambda_n}}^{+}+\beta_{k_{\lambda_n}}^{-})}_{{\text{penalty term}}}}}\nonumber
\end{IEEEeqnarray} 
\begin{IEEEeqnarray}{lCr}
\scaleobj{0.8}{\text{s.t.}\hspace{1em} \B{\beta_{\lambda_n}^{+}}\ge \B{0}, \hspace{1em} \B{\beta_{\lambda_n}^{-}}\ge \B{0}} \nonumber
\end{IEEEeqnarray}
where the modified solution path is $\B{\varphi}_{\lambda_n}^*=\LeftCurly{1.2}{\B{\theta_{\lambda_n},\B{\beta_{\lambda_n}^{+}}}}$, $\RightCurly{1.2}{\B{\beta_{\lambda_n}^{-}},\B{\eta_{_{\lambda_n}}},\B{\delta_{\lambda_n}},\B{\alpha_{_{\lambda_n}}}}$.

Note that the the sum of squares term in \eqref{Problem:convex} is unaffected, if we set $\B{\beta_{\lambda_n}^{+}}\longleftarrow\B{\beta_{\lambda_n}^{+}}+\B{s}$ and $\B{\beta_{\lambda_n}^{-}}\longleftarrow\B{\beta_{\lambda_n}^{-}}+\B{s}$ $\forall\B{s}\ge \B{0}$, because $\B{s}$  is canceled out in \eqref{Difference:Operator}.\footnote{$\B{s}$ cannot contain negative elements, because the set $(\beta_{k_{\lambda_n}}^{+},\beta_{k_{\lambda_n}}^{-})=(\beta_{k_{\lambda_n}},0)$ implies $\beta_{k_{\lambda_n}}>0$, while the set  $(\beta_{k_{\lambda_n}}^{+},\beta_{k_{\lambda_n}}^{-})=(0,-\beta_{k_{\lambda_n}})$ implies $\beta_{k_{\lambda_n}}<0$. The intuition being that if $\B{s}<0$, the requirements $\B{\beta_{\lambda_n}^{+}}+\B{s}\ge 0$ and $\B{\beta_{\lambda_n}^{-}}+\B{s}\ge 0$ are not satisfied. If $\B{s}>0$ the penalty function is not minimized. } However, the argument in the penalty function term increases by $2\B{s}$. As a result, $\B{s}=0$ minimizes the penalty function, implying that the solution of problem \eqref{Problem:convex} for a given $k$ is either $\beta_{k_{\lambda_n}}^{+}=0$ or $\beta_{k_{\lambda_n}}^{-}=0$. Problem \eqref{Problem:convex} is equivalent to the original problem \eqref{SPLM:SCAD}, where
$\abs{\beta_{k_{\lambda_n}}}=\beta_{k_{\lambda_n}}^{+}+\beta_{k_{\lambda_n}}^{-}$ and $\beta_{k_{\lambda_n}}=\beta_{k_{\lambda_n}}^{+}-\beta_{k_{\lambda_n}}^{-}$ $\forall k$.
The aforementioned argument points to the possibility of using simple constrained convex penalty function algorithms for the estimation of the optimal number of reference groups; this is embedded in the selection equation, which is affected by the number of reference groups. Technical details appear in Appendix \ref{Section:Penalty:function:Decomposition:Appendix}.

\section{Simulations}\label{Sec:Simulations}

We examine our truncated selection model's performance in the presence of various latent classes, capturing the unobserved characteristics of each datum. A sequence $\curly{(\Lambda_{k}^S,\Lambda_{k}^T)}_{k=1}^{10,000}$ consisting of $10,000$ elements is generated. The $k$'th element is composed of a survey data set and a truncated data set denoted by $\Lambda_k^S$  and $\Lambda_k^T$, respectively. For simplicity, the data sets are generated using three latent classes.  

Next we discuss the data generation process (DGP) used to construct these distribution functions.

\subsection{Data Generation Process (DGP)}\label{DGP}
Let $\mathrm{z}$ be a continuous random variable, such that a given realization of this random variable represents specific individual level characteristics.  Our objective is to characterize a sequence of distribution functions $\curly{\mathcal{D}_{\mathrm{z}|\B{x}_g}}_{g=1}^{G}$. Each is a conditional distribution function of $\mathrm{z}$, given a specific realization of the group-level observed characteristics, $\B{x}_g$. These distribution functions are not restricted to being unimodal or symmetric (e.g., the normal distribution function).\footnote{Unlike the Monte Carlo simulations in \cite{breunig2017nonparametric} for censored sample selection models implemented by using normally distributed disturbances, we consider a truncated sample selection model characterized by non-normally distributed disturbances.} By employing such an algorithm, each datum in the truncated data set to be generated is a random draw from its group-specific  distribution function. We arbitrarily set $G=2,000,000$ indicating the number of distribution functions in the sequence. The conditional density function of $\mathrm{z}$ given $\B{x}_g$ is denoted by $\mathlcal{d}_{z|\B{x}_g}(z|\B{x}_g)$ and satisfies $\forall t=1,...,\tilde{G}$:
\begin{IEEEeqnarray}{lCr}\label{Equilibria:DGP}
m(\B{x}_g,t)=\int \LeftClose{1.8}{1-F_{\disturb_{2}}(\alpha+\beta m(\B{x}_g,t)}\\
	+\RightClose{1.8}{\close{\B{x}_g^c}^T\delta  + \eta z)}\mathlcal{d}_{z|\B{x}_g}(z|\B{x}_g) dz,\hspace{1em} t=1,...,\tilde{G}\nonumber
\end{IEEEeqnarray}
where $m(\B{x}_g,t)$ is the actual participants' share given $\B{x}_g$ and being a member in class $t$, and $\B{x}_g^c$ is a subset of $\B{x}_g$, consisting of the contextual covariates only.

Finding a density function $\mathlcal{d}_{z|\B{x}_g}(z|\B{x}_g)$ that satisfies \eqref{Equilibria:DGP} is computationally cumbersome, due to the presence of the integral. In order to facilitate the computation process, this density is expressed as a finite mixture of arbitrarily chosen continuous density functions, such that only the weights (mixture coefficients) are required to uncover.  

Thus, let $\curly{\phi_l(.): l=1,...,L}$ be an arbitrary set of distinct continuous
probability density functions on the real line\footnote{We utilize a mixture consisting of various density functions, including normal, gamma and log-normal. Each density function is characterized by a unique set of parameters.} and $\curly{\Phi_l(.): l=1,...,L}$ be the corresponding distribution functions. The mixture of the probability density
functions using the
weights $w_l$, satisfying $\sum_{l=1}^{L}w_l=1$, is defined as follows: 
\begin{IEEEeqnarray}{lCr}\label{Mixture:densities}
\mathlcal{d}_{z|\B{x}_g}(z|\B{x}_g) = \sum_{l=1}^{L}w_l(\B{x}_g)\phi_l(z) 
\end{IEEEeqnarray}
and the mixture of the distribution functions is
\begin{IEEEeqnarray}{lCr}\label{Mixture:Distributions}
\mathcal{D}_{z|\B{x}_g}(z|\B{x}_g) = \sum_{l=1}^{L}w_l(\B{x}_g)\Phi_l(z) 
\end{IEEEeqnarray}

Denote $\mathfrak{q}(p,z)\equiv 1-F_{\disturb_{2}}(\alpha+\beta p
+\close{\B{x}_g^c}^T\delta  + \eta z)$. The sequence of optimal weights, $\curly{w_l}_{l=1}^L$, consists of $L$ elements such that $\sum_{l=1}^{L}w_l=1$ and $0\le w_l \le 1$. The following condition must hold for all $t=1,...,\tilde{G}$:
\begin{IEEEeqnarray}{lCr}\label{Distribution:Function:Matching}
\sum_{l=1}^{L} \braket{\int\mathfrak{q}\close{m(\B{x}_g,t),z}\phi_l(z) dz}w_l=m(\B{x}_g,t), \hspace{1em} 
\end{IEEEeqnarray}

A matrix $\mathbf{M}_{\scaleto{\mathcal{P}}{4pt}}$ of size $\tilde{G}\times L$ and a vector $\mathbf{w}$ of size $L\times 1$ are defined as:
\begin{IEEEeqnarray}{lCr}\label{Define:M}
\mathbf{M}_{\scaleto{\mathcal{P}}{4pt}}=\scaleobj{0.8}{\braket{\begin{matrix}
	\int\mathfrak{q}(m(\B{x}_g,1),z)\phi_1(z) dz &...& \int\mathfrak{q}(m(\B{x}_g,1),z)\phi_L(z) dz \\
	: & \ddots & : \\
	\int\mathfrak{q}(m(\B{x}_g,\tilde{G}),z)\phi_1(z) dz &...& \int\mathfrak{q}(m(\B{x}_g,\tilde{G}),z)\phi_L(z) dz 
	\end{matrix}}_{\tilde{G}\times L}} \hspace{1em}\nonumber\\ \scaleobj{0.8}{\mathbf{w}=\braket{
	\begin{matrix} 
	w_1(\B{x})\\
	w_2(\B{x})\\
	:\\
	w_L(\B{x})
	\end{matrix}}_{L\times 1}}\nonumber
\end{IEEEeqnarray}

These optimal weights, which solve \eqref{Distribution:Function:Matching}, can be obtained as a solution to the following minimization problem:
\begin{IEEEeqnarray}{lCr}\label{Discritize:Distribution}
\mathbf{w}=\underset{\substack{\text{s.t.}\hspace{1em}0\le w_h\le 1 \\ \\ \hspace{2em}\sum_{h}^{H}w_h=1  \\ \\ \hspace{2em}\mathbf{M_{\scaleto{\mathcal{P}'}{4pt}}w}>\mathcal{P}'  \\ \\ \hspace{2em}\mathbf{M_{\scaleto{\mathcal{P}''}{4pt}}w}<\mathcal{P}''}}{\arg\min}\normsq{\mathbf{M_{\scaleto{\mathcal{P}}{4pt}}w}-\mathcal{P}}{2}
\end{IEEEeqnarray}
where $\mathcal{P}=[m(\B{x}_g,1),...,m(\B{x}_g,\tilde{G})]^T$  is a $\tilde{G}\times 1$ vector, $\mathbf{M_{\scaleto{\mathcal{P}'}{4pt}}}$ and  $\mathbf{M_{\scaleto{\mathcal{P}''}{4pt}}}$ are matrices constructed in a similar fashion to $\mathbf{M_{\scaleto{\mathcal{P}}{4pt}}}$ (described in \eqref{Define:M}) to ensure that the expected participants' shares vector, $\mathcal{P}$, is the unique solution for \eqref{Discritize:Distribution} such that any other expected participant share depicted in either vector $\mathcal{P}'$ of size $J'\times 1$  or vector $\mathcal{P}''$ of size $J''\times 1$   will not constitute a solution. 

The main idea is that for any vector $\mathcal{P}$ consisting of participants' shares (calculated for a given observed group's characteristics), we match a distribution function of  individual characteristics, such that \eqref{Equilibria:DGP} is satisfied.\footnote{By construction, the generated distribution function leads to a unique solution characterized by vector $\mathcal{P}$.} Formally, the $k$'th data consisting of $N$ observations is generated by using a sequence $\curly{\B{x_i}}_{i=1}^N$  of randomly drawn vectors from a joint distribution function $F_{\mathrm{\B{x}}}$ (as will be described in \eqref{Joint:CDF:X} to follow) and a sequence of distribution functions $\curly{\mathcal{D}_{z|\B{x_i}}(z|\B{x_i})}_{i=1}^N$. Each data point $\B{z_i}$ of individual level characteristics is a realization of a unique random variable $\B{\mathrm{z|x}=x_i}$, drawn from $\mathcal{D}_{z|\B{x_i}}(z|\B{x_i})$, where $\B{x}_i=[x_{i}, x_{i}^c]$ and represents the observed group's characteristics of the $i$'th observation.\footnote{$z_i$ is a single covariate, so we use a univariate distribution function. However, in cases where it is a covariate vector, the index function $\B{z}_i^T\B{\eta}$ can be treated as a random realization from a univariate distribution function $\mathcal{D}_{\B{z}_i^T\B{\eta}|\B{x_i}}(\B{z}_i^T\B{\eta}|\B{x_i})$.}

Based on \eqref{Equilibria:DGP}, each of the distribution functions to be found is required to satisfy a restriction concerning a specific participant shares vector. Therefore, these shares must be known in the data generation process.\footnote{Thus, we define the $t$'th element of $\mathcal{P}(\B{x}_g)$ as:
	\begin{IEEEeqnarray}{lCr}\label{Participation:Share}
	\scaleobj{0.7}{m^{\mathlcal{e}}(\B{x}_g,t)=\mathlcal{p}_{lt}+(\mathlcal{p}_{ht}-\mathlcal{p}_{lt})\frac{\exp\close{-2+3.6 \close{\sqrt{\Phi_{_{\mathcal{N}}}(x_{g})\Phi_{_{\mathcal{N}}}(x_{g}^c)}}}}{1+\exp\close{-2+3.6 \close{\sqrt{\Phi_{_{\mathcal{N}}}(x_{g})\Phi_{_{\mathcal{N}}}(x_{g}^c)}}}}}\nonumber
	\end{IEEEeqnarray}
	where $\close{\mathlcal{p}_{lt}, \mathlcal{p}_{ht}}$ is the range of the conditional participation probability given a membership in class $t$. The ranges are arbitrarily determined to be: $\close{\mathlcal{p}_{l1}, \mathlcal{p}_{h1}}=\close{0.05, 0.4}$, $\close{\mathlcal{p}_{l2}, \mathlcal{p}_{h2}}=\close{0.4, 0.75}$ and $\close{\mathlcal{p}_{l3}, \mathlcal{p}_{h3}}=\close{0.65, 0.95}$. These numbers enable us to verify the model performance in cases where the participants' share is a non-smooth function of the group's observed characteristics ($\B{x}_g$). This non-smoothness stems from the presence of latent classes, which are determined as a function of $\B{x}_g$ (due to non-random assignment).  The cumulative standard normal distribution function is denoted by $\Phi_{_{\mathcal{N}}}$.	 }

\subsection{Generation of survey and truncated data sets utilizing latent classes}\label{Survey data}
The group's characteristics covariate vector $\B{x}_g=[x_{g}, x_{g}^c]$ is a realization of the random variables vector $\B{\mathrm{x}}=[\mathrm{x,x^c}]$, which is jointly distributed $F_{x}$. For simplicity we characterize $F_x$ as follows:
\begin{IEEEeqnarray}{lCr}\label{Joint:CDF:X}
[\mathrm{x,x^c}]\sim\mathcal{N}_2\close{\braket{\begin{matrix}
		0 \\ 0
		\end{matrix}}, \braket{\begin{matrix}
		2 & 0.5\sqrt{2*3}\\ 0.5\sqrt{2*3} & 3
		\end{matrix}}}
\end{IEEEeqnarray}
where $\mathcal{N}_2$ denotes the bivariate normal distribution function.

The classes and a frequency table consisting of manifest variables (depicted in Table \ref{tabSurvey} to follow) are generated in $R$, using the latent classes packages $`poLCA'$ and $`SimCorMultRes'$. For each $k=1,...,10,000$, two data sets are generated: (i) a survey data set $\Lambda_k^S=\curly{\varphi_i^*,\chi_{\varphi_i}}_{i=1}^{N^E}$ with $\varphi_i=(\varphi_i^*,\psi_i^E)\in\Lambda$ (constructed as depicted in section \ref{Sec:Opinions:Experts}) consisting of a sequence $\curly{\varphi_i^*}_{i=1}^{N^E}$, in which $\varphi_i^*=(\SizeSym{X}_i,\SizeSym{Y}_i)$ are the experts' observed characteristics and $N^E=10,000$; (ii) a truncated data set $\Lambda_k^T=\curly{(\B{x},\SizeSymGen{Y},y_{1}^*,z)|(\B{x},\SizeSymGen{Y},y_{1}^*,y_{2}^*,z,\psi)\in \Lambda_k^C,y_{2}^*\ge0}$, where $\Lambda_k^C=\curly{\B{x}_i,\SizeSymGen{Y}_{i},y_{1i}^*,y_{2i}^*,z_i,\psi_i}_{i=1}^{N}$ denotes the complete (non-truncated) data set. The number of observations in the truncated data set is denoted by $n_{k}$, which is the cardinality of the set $\Lambda_k^T$.\footnote{The truncated data set is produced by generating a complete (non-truncated) data set and keeping only the observations that satisfies the selection equation.} The set $\Lambda_k^T$ consists of classes, manifest variables and observed group characteristics $\B{x}$ randomly and independently drawn from \eqref{Joint:CDF:X}. These characterize the group's observed characteristics and individual-level covariates, as denoted by the sequence $\curly{z_i}_{i=1}^n$. Using the estimated posteriori classes density function and the survey data set, the predicted class, $\widehat{\psi}_{i,g}$, for each observation $i$ is calculated in the truncated data set (given the manifest variables and the group's observed characteristics\footnote{The observed group's characteristics affect the prior class membership assignment probabilities.}).  These predicted classes are intended to be used later, in the estimation stage, and not in the data generation process. The selection model's equation will be generated by using the true classes, as will be described in section \ref{Sec:Equations:Generation}.

The characteristics $\SizeSym{X}_i=[x_i^E,(x_i^c)^E]$  are randomly and independently drawn from \eqref{Joint:CDF:X}. The $i$'th observation's latent class is generated as a random realization from \eqref{Prior:Probability:Function}, which is a function of  $\SizeSym{X}_i$.\footnote{Due to the non-random assignment, the $i$'th observation's latent class membership depends on $\SizeSym{X}_i$.} The manifest variables are determined using the frequencies in Table \ref{tabSurvey} and are generated for each $i\in{1,...,N^E}$.

Table \ref{tabSurvey} exhibits the construction of the manifest categorical variables in the survey data set, which are denoted by $\mathlcal{y}_{_1}$,...,$\mathlcal{y}_{_7}$. These variables enable the recovery of the latent classes. Recovery is feasible due to the conditional independence (given the class) property. It follows that the probability of observing a specific response (the alternative being chosen) in $\mathlcal{y}_{_k}$ is independent of the response to $\mathlcal{y}_{_l}$ for all $k\ne l$, given the class membership.

\begin{table}
	
	\caption{\label{tabSurvey}Response frequencies for the manifest categorical variables}
	
	\begin{center}\scalebox{1}[0.7]{\begin{adjustbox}{width=25em}\begin{tabular}{|c|cccc|cccc|cccc|}\hline\hline 
				\multicolumn{1}{|c|}{\multirow{3}{*}{Manifest variable}} & \multicolumn{4}{c|}{Class 1} & \multicolumn{4}{c|}{Class 2} & \multicolumn{4}{c|}{Class 3}\\ \cline{2-13} 
				\multicolumn{1}{|c|}{} & \multicolumn{4}{c|}{Response} & \multicolumn{4}{c|}{Response}  & \multicolumn{4}{c|}{Response}\\ \cline{2-13} 
				\multicolumn{1}{|c|}{} & \multicolumn{1}{c|}{1} & \multicolumn{1}{c|}{2} & \multicolumn{1}{c|}{3} & \multicolumn{1}{c|}{4} & \multicolumn{1}{c|}{1} & \multicolumn{1}{c|}{2} & \multicolumn{1}{c|}{3} & \multicolumn{1}{c|}{4} & \multicolumn{1}{c|}{1} & \multicolumn{1}{c|}{2} & \multicolumn{1}{c|}{3} & \multicolumn{1}{c|}{4} \\ \hline 
				${\mathlcal{y}_1}$ & 0.6 & 0.1 & 0.3 & - & 0.4 & 0.4 & 0.2 & - & 0.3 & 0.1 & 0.6 & - \\\hline 
				${\mathlcal{y}_2}$ & 0.2 & 0.8 & - & - & 0.5 & 0.5 & - & - & 0.7 & 0.3 & - & - \\\hline
				${\mathlcal{y}_3}$ & 0.3 & 0.6 & 0.1 & - & 0.1 & 0.4 & 0.5 &  & 0.7 & 0.2 & 0.1 &  \\\hline
				${\mathlcal{y}_4}$ & 0.1 & 0.6 & 0.2 & 0.1 & 0.5 & 0.3 & 0.1 & 0.1 & 0.3 & 0.1 & 0.1 & 0.5 \\\hline
				${\mathlcal{y}_5}$ & 0.1 & 0.1 & 0.8 & - & 0.6 & 0.3 & 0.1 & - & 0.8 & 0.1 & 0.1 & - \\\hline
				${\mathlcal{y}_6}$ & 0.9 & 0.02 & 0.08 & - & 0.02 & 0.08 & 0.9 & - & 0.08 & 0.9 & 0.02 & - \\\hline
				${\mathlcal{y}_7}$ & 0.95 & 0.05 & - & - & 0.05 & 0.95 & - & - & 0.5 & 0.5 & - & - \\\hline

	\end{tabular} \end{adjustbox} }\end{center} 
	\scriptsize\begin{minipage}{8.8cm}%
		$\myspace$ \\ \textbf{Note:}  The survey data set consists of the manifest categorical variables set $\mathlcal{y_{_1}},...,\mathlcal{y_{_7}}$: variable $\mathlcal{y_{_4}}$ is of four categories (alternative responses); each of the variables $\mathlcal{y_{_1}},\mathlcal{y_{_3}},\mathlcal{y_{_5}},\mathlcal{y_{_6}}$ are of three categories; the variables  $\mathlcal{y_{_2}}$ and $\mathlcal{y_{_7}}$ are binary response variables. All of these manifest variables are conditionally independent, with respect to each other, given the class membership. It follows that a membership in class $'1'$ implies that the probability to choose response $'2'$ in $\mathlcal{y_{_1}}$ is $0.1$, regardless of all other responses.
	\end{minipage}		
	\footnotesize
	\renewcommand{\baselineskip}{11pt}		
\end{table} 

Next we construct the selection model's joint disturbances distribution function, in order to examine our model's performance in cases of a non-standard distribution function of the disturbances (such as the normal distribution function).  
\subsection{The disturbances' joint distribution function}

Each pair of disturbances $\curly{\disturb_{1i},\disturb_{2i}}$ is randomly and independently drawn from $F_{\disturb_1,\disturb_2}$, which is the joint distribution function of the substantive and participation equations' disturbances.  The aforementioned joint density function consists of two components: a Copula function\footnote{According to Sklar's Theorem \cite{sklar1959fonctions}, any continuous joint distribution function can be characterized by a set of marginal distribution functions and a joint distribution function determining the dependence structure which is referred to as a Copula function.} characterizing the disturbances' dependence structure and two marginal distribution functions $F_{\disturb_1}$ and $F_{\disturb_2}$ for the substantive equation and selection equation, respectively. In order to verify our model's performance in the presence of random disturbances' distribution functions which are not restricted to the family of symmetric and unimodal distribution functions, each one of these two disturbances is marginally distributed according to a mixture of three different distribution functions: (i) a normal distribution function with expectation and standard deviation parameters $(\mu, \sigma_a)$, denoted by $\mathcal{N}(\mu,\sigma_{a}^2)$; (ii) a  normal distribution function with expectation and standard deviation parameters $(-\mu, \sigma_b)$, denoted by $\mathcal{N}(-\mu,\sigma_{b}^2)$; (iii) a gamma distribution function with scale and shape parameters $(\mu\varphi,\varphi)$, denoted by $\Gamma_{\scaleto{\mathcal{\text{Gamma}}}{4pt}}\close{\mu\varphi, \varphi}$\footnote{The scale and shape parameters implies that the expectation and standard deviation parameters are $(\mu ,\sqrt{\mu/\varphi})$.}. This mixture distribution function is defined as: 
\begin{IEEEeqnarray}{lCr}
\\ \disturb_{j}\sim 0.4\mathcal{N}(\mu,\sigma_{a}^2)+0.5\mathcal{N}(-\mu,\sigma_{b}^2)+0.1\Gamma_{\scaleto{\mathcal{\text{Gamma}}}{4pt}}\close{\mu\varphi, \varphi}\nonumber
\end{IEEEeqnarray}
where $\mathbb{E}\braket{\disturb_{j}}=0$, $j=1,2$.

The parameters set $(\mu,\sigma_a,\sigma_b,\varphi)=(4,3.5,2.5,2)$ is arbitrarily chosen. Due to its simplicity, the Clayton Copula, with a degree of dependence parameter $4$ (to assure that the disturbances are highly correlated) is used for controlling the dependence structure.

\subsection{Constructing the selection model's equations}\label{Sec:Equations:Generation}\label{Sec:Simulation:Equations}
In the last step, we construct the latent selection equation's dependent variable $y_{2i,g}^*$ for $i=1,...,N$:
\begin{IEEEeqnarray}{lCr}
y_{2i,g}^*=\alpha+\beta m(\B{x}_{g},\psi_{i,g})+\B{z}_i^T\B{\eta}+(x_{g}^c)^T\delta+\disturb_{2i},
\end{IEEEeqnarray}
where the $i$'th observation's true class membership is denoted by a categorical variable\\ $\psi_{i,g}\in\curly{1,2,3}$ and $\B{x}_g=[x_g,x_g^c]$. 

In a similar fashion, we construct the substantive equation's dependent variable:
\begin{IEEEeqnarray}{lCr}
y_{1i,g}^*=\theta_1 w_i + \theta_2 D_i +\disturb_{1i},\myspace i=1,...,N.
\end{IEEEeqnarray}
where any covariates pair  $(w_i, D_i)$ is an independent  realization of a random variable vector $(\mathrm{w,D})|z_i$, conditionally distributed given $z_i$, such that $\mathrm{w}|z_i$ and  $\mathrm{D}|z_i$ are independent.\footnote{This assumption is termed a conditional independence given $z_i$ and is imposed to reduce the complexity of the data generation process.} These conditional random variables are distributed according to a normal and a Bernoulli probability distribution function, respectively. 
We arbitrarily set $(\alpha,\beta,\delta)=(-30, 25, 1.5)$ and $(\theta_1, \theta_2)=(2,4)$.

Denote the selection variable by an indicator function  $y_{2i,g}=I(y_{2i,g}^*\ge 0)$ or alternatively, calculate the $i$'th observation's probability of being selected $p_i=\Pr(\Bigger{\omega}_{i,g}=1|x_{g},x_{g}^c,\psi_{i,g},z_i)$, which is a function of $(x_{g},x_{g}^c,\psi_{i,g},z_i)$ and defined for $i=1,...,N$ as:
\begin{IEEEeqnarray}{lCr}
p_{i,g}=1-F_{\disturb_2}(\alpha+\beta m(\B{x}_{g},\psi_{i,g})+\B{z}_i^T\B{\eta}+(x_{g}^c)^T\delta)\hspace{1em}
\end{IEEEeqnarray}

Let $\curly{u_1,...,u_N}$ be a sequence of continuous and independent uniform random variables on the support $[0,1]$. The indicator variable for the $i$'th observation is:
\begin{IEEEeqnarray}{lCr}
y_{2i,g}=I(u_i\le p_i),
\end{IEEEeqnarray}
that is, $y_{2i,g}$ is the realization of a Bernoulli distributed random variable with a probability of success, $p_{i,g}$, with $p_{i,g}$ the probability of observation $i$, in reference group $g$, to be observed in the truncated data.

The final truncated data set consists of two sequences of the self-selected observation (satisfying $y_{2i}=1$).  The truncated data set consists of 50$\%$ of the observations.\footnote{Following  \cite{arabmazar1982investigation}, employing a parametric (censored or truncated) sample selection model and misspecifying the random disturbances to be joint normal distributed might lead to bias in the estimates, such that its magnitude depends on the degree of censoring or truncation of the sample. They find evidence that the bias is substantial, especially for truncated samples that are 50 percent complete. Because our model is distribution-free, it is important to verify its performance given those conditions in which the parametric models underperformed. Thus, we use a truncated data set which is 50$\%$ complete.} The first includes the individual level covariates $\curly{y_{1i}^*,z_i,w_i,D_i|y_{2i}=1}_{i=1}^N$, and the second includes the group's observed characteristics and the manifest categorical variables:  $\curly{x_{1i},x_{2i},\mathlcal{y}_{1i},..,\mathlcal{y}_{7i}|y_{2i}=1}_{i=1}^N$. Similarly, the survey data set consists of the sequence $\curly{\bar{\mathcal{P}},x_{1i}^{\text{survey}},x_{2i}^{\text{survey}},\mathlcal{y}_{1i}^{\text{survey}},..,\mathlcal{y}_{7i}^{\text{survey}}}_{i=1}^N$.\footnote{Of course, the true class sequence in both the truncated data set $\curly{c_i}_{i=1}^N$ as well as the survey data set $\curly{c_i^\text{survey}}_{i=1}^N$ is unobserved. Thus, they are excluded from these data sets.} The aggregate participants' share in the entire population is denoted by $\bar{\mathcal{P}}$.

The main results regarding the estimates obtained using $10,000$ Monte Carlo simulations for sample sizes $N\in\curly{2000,5000,8000,10000}$, as depicted in sections \ref{DGP}-\ref{Sec:Simulation:Equations}, are summarized in Tables \ref{tabB} and \ref{tabA} to follow.

\begin{table}
	\caption{\label{tabB}Monte Carlo Simulation - OLS without truncation bias correction$^{\mathrm{a}}$}
	
	\begin{center}\scalebox{0.9}{\begin{adjustbox}{width=25em}\begin{tabular}{ccccccccc|cccc|}\hline\hline 
				\multicolumn{3}{|c|}{\multirow{4}{*}{True Parameter$^{\mathrm{b}}$}} & \multicolumn{2}{l|}{\multirow{4}{*}{Estimate}} & \multicolumn{8}{c|}{Model Type}\\ \cline{6-13} \multicolumn{3}{|l|}{}  & \multicolumn{2}{l|}{} & \multicolumn{4}{c|}{Full sample} & \multicolumn{4}{c|}{Truncated sample}\\ \cline{6-13}   
				\multicolumn{3}{|l|}{}  & \multicolumn{2}{l|}{} & \multicolumn{4}{c|}{Sample size} & \multicolumn{4}{c|}{Sample size}  \\ \cline{6-13} 
				
				\multicolumn{3}{|l|}{}  & \multicolumn{2}{l|}{}  & \multicolumn{1}{c|}{2000} & \multicolumn{1}{c|}{5000} & \multicolumn{1}{c|}{8000} & \multicolumn{1}{c|}{10000} & \multicolumn{1}{c|}{2000} & \multicolumn{1}{c|}{5000} & \multicolumn{1}{c|}{8000} & \multicolumn{1}{c|}{10000} \\ \hline  \hline
				
				\multicolumn{3}{|c|}{\multirow{3}{*}{$\theta_1$=2}}     & \multicolumn{2}{c|}{Mean}   & 1.9989 & 1.9990 & 1.9995 & 2.0002 & 2.5873 & 2.5858 & 2.5853 & 2.5851 \\ \cline{4-13} \multicolumn{3}{|c|}{} & \multicolumn{2}{c|}{Median}  &  1.9986 & 1.9987 & 2.0002 &  1.9999 &  2.5883 & 2.5859 & 2.5852 & 2.5851 \\ \cline{4-13} \multicolumn{3}{|c|}{} & \multicolumn{2}{c|}{Std}  & 0.0595 & 0.0373 & 0.0294 & 0.0265 & 0.0767 & 0.0481 & 0.0380 & 0.0350\\ \hline \hline
				
				\multicolumn{3}{|c|}{\multirow{3}{*}{$\theta_2$=4}} & \multicolumn{2}{c|}{Mean} & 4.0110 & 4.0076 & 4.0051 & 3.9984 & 4.2916 & 4.2897 & 4.2838 & 4.2587 \\ \cline{4-13} \multicolumn{3}{|c|}{} & \multicolumn{2}{c|}{Median}  &  4.0140 & 4.0055 & 4.0048 & 4.0012 & 4.2862 & 4.2842 & 4.2806 & 4.2591 \\ \cline{4-13} \multicolumn{3}{|c|}{} & \multicolumn{2}{c|}{Std}  & 0.5834 & 0.3705 & 0.2922  & 0.2583 & 0.9917 & 0.6279 & 0.4931 & 0.4492\\\hline							
				
	\end{tabular}\end{adjustbox} }\end{center}
	\scriptsize\begin{minipage}{8.2cm}%
		$\myspace$ \\ \textbf{Note:}  $^{\mathrm{a}}$ We estimate by ordinary least squares (OLS) method the parameters for the full sample and truncated sample without correction for the selectivity bias, and compute the standard deviation in every random sample consisting of N observations. Then, we calculate for these estimates the mean, median and standard deviation (Std.) over all data sets. The standard deviations are obtained using the estimates from the Monte-Carlo simulations. \\  $^{\mathrm{b}}$ The parameters that are used in the data generation process. $\theta_1$ and $\theta_2$ are the parameters of interest in the substantive equation. 
	\end{minipage}		
	\footnotesize
	\renewcommand{\baselineskip}{11pt}		
\end{table}

The entries in Table \ref{tabB} indicate that for a sample size of $2000$ observations, the mean estimate of $\theta_1$ in the full sample is $2.000$, while in the truncated sample, without correction for bias is $2.5873$. Similarly, for the same sample size, the mean estimate of $\theta_2$ in the full sample is $4.000$, while in the truncated sample without correction for bias is $4.2916$. The estimates for $\theta_1$ and $\theta_2$ are remained biased (upward) as the sample size increases.

\newlength\q
\setlength\q{\dimexpr .1\textwidth -2\tabcolsep}
\newlength\qq
\setlength\qq{\dimexpr .12\textwidth -2\tabcolsep}

\begin{table}
	\caption{\label{tabA}Monte Carlo Simulation - NLS$^{\mathrm{a}}$ with truncation bias correction$^{\mathrm{b}}$}
	
	\begin{center}\scalebox{0.9}{\begin{adjustbox}{width=25em}\begin{tabular}{ccccccccc|cccc|}\hline\hline 
				\multicolumn{3}{|c|}{\multirow{4}{*}{True Parameter$^{\mathrm{c}}$}} & \multicolumn{2}{l|}{\multirow{4}{*}{Estimate}} & \multicolumn{8}{c|}{Model Type}\\ \cline{6-13} \multicolumn{3}{|l|}{}  & \multicolumn{2}{l|}{} & \multicolumn{4}{c|}{Refined Vox Populi} & \multicolumn{4}{c|}{Monolithic Vox Populi}\\ \cline{6-13}   
				\multicolumn{3}{|l|}{}  & \multicolumn{2}{l|}{} & \multicolumn{4}{c|}{Sample size} & \multicolumn{4}{c|}{Sample size}  \\ \cline{6-13} 
				
				\multicolumn{3}{|l|}{}  & \multicolumn{2}{l|}{}  & \multicolumn{1}{c|}{2000} & \multicolumn{1}{c|}{5000} & \multicolumn{1}{c|}{8000} & \multicolumn{1}{c|}{10000} & \multicolumn{1}{c|}{2000} & \multicolumn{1}{c|}{5000} & \multicolumn{1}{c|}{8000} & \multicolumn{1}{c|}{10000} \\ \hline  \hline
				
				
				\multicolumn{3}{|c|}{\multirow{3}{*}{$\theta_1$=2}} & \multicolumn{2}{c|}{Mean} & 2.0083 & 2.0080 & 2.0076 &  2.0070 &  2.0509 & 2.0505 & 2.0483 & 2.0388   \\ \cline{4-13} \multicolumn{3}{|c|}{} & \multicolumn{2}{c|}{Median}   & 2.0085 & 2.0077 & 2.0069 & 2.0063 & 2.0517 & 2.0497 & 2.0494 & 2.0374
				\\ \cline{4-13} \multicolumn{3}{|c|}{} & \multicolumn{2}{c|}{Std} & 0.2424 & 0.1535 & 0.1207 & 0.1075 & 0.2428 & 0.1542 & 0.1211 & 0.1086\\ \hline \hline
				
				\multicolumn{3}{|c|}{\multirow{3}{*}{$\theta_2$=4}} & \multicolumn{2}{c|}{Mean} & 3.9909 & 3.9954 & 3.9971 & 3.9978 & 3.9072 & 3.9127 & 3.9151 & 3.9157 \\ \cline{4-13} \multicolumn{3}{|c|}{} & \multicolumn{2}{c|}{Median}  &  3.9924 & 3.9943 & 4.0035 & 4.0003 & 3.9083 & 3.9145 & 3.9159 & 3.9171
				\\ \cline{4-13} \multicolumn{3}{|c|}{} & \multicolumn{2}{c|}{Std}  & 1.012 & 0.6261 & 0.4910 & 0.4426 & 1.0105 & 0.6256 & 0.4914 & 0.4435\\\hline

	\end{tabular}\end{adjustbox} }\end{center}
	\scriptsize\begin{minipage}{8.2cm}%
		$\myspace$ \\ \textbf{Note:} $^{\mathrm{a}}$Nonlinear Least Squares.\\ $^{\mathrm{b}}$ We estimate by a semiparametric nonlinear least squares (NLS) method the parameters for the truncated sample under refined and monolithic Vox-Populi specifications. In the former specification, the model is estimated using the true number of latent classes (treated as given), while in the latter there are no latent classes. The bias term function, the nonlinear part of the regression, is approximated using a transformed Fourier series as depicted in section \ref{Sec:Esitmate:Fourier}. Then, we calculate for these estimates the mean, median and standard deviation (Std.) over all data sets. The standard deviations are obtained using the estimates from the Monte Carlo simulations.\\ $^{\mathrm{c}}$ The parameters that are used in the data generation process. $\theta_1$ and $\theta_2$ are the parameters of interest in the substantive equation.  
	\end{minipage}		
	\footnotesize
	\renewcommand{\baselineskip}{11pt}	 
\end{table}

Entries presented in Table \ref{tabA} indicate that the substantive equation's estimated parameters, $\widehat{\theta}_1=2.0083$ and $\widehat{\theta}_2=3.9909$, in the refined Vox Populi model (given the correct number of latent classes) almost mimic the parameters estimates that would have been generated from a non-truncated sample, which are $\theta_1=2.000$ and $\theta_2=4.000$ respectively, for a sample size of 2,000 observations.\footnote{The parameter estimates start deteriorating below $2000$ observations in both the penalized and the non-penalized refined Vox Populi models, and thus are not presented in the table.} However,  treating the truncated data as if it consist of a single class (a monolithic Vox Populi) in the presence of multiple latent classes, the substantive equation's parameters' estimates are $\widehat{\theta}_1=2.0509$ and $\widehat{\theta}_2=3.9083$, given a sample of 2,000 observations. The standard deviation for each estimated parameter is calculated over the estimates obtained from all the Monte Carlo simulations,\footnote{The standard deviations are calculated using the same methodology for each of the models, to make it easier to compare results from different regression models.} for each one of the refined and monolithic Vox Populi models. The means of the parameter estimates obtained for $\theta_1$ and $\theta_2$ in the refined Vox Populi model are  $2.007$ and $3.9978$, respectively, for a sample size of 10,000 observations. While in the monolithic Vox Populi model, the means of the parameter estimates obtained for $\theta_1$ and $\theta_2$ in the selection equation, are  $2.0388$ and $3.9157$, respectively, using the same sample size.\footnote{The nuisance parameters' estimates, consisting of the selection equation's estimated coefficients, will be supplied upon request.}

\begin{table}
	\caption{\label{tabC}Monte Carlo Simulation - NLS$^{\mathrm{a}}$ with refined Vox Populi truncation bias correction$^{\mathrm{b}}$}
	
	\begin{center}\scalebox{0.9}{\begin{adjustbox}{width=25em}\begin{tabular}{ccccccccc|cccc|}\hline\hline 
				\multicolumn{3}{|c|}{\multirow{4}{*}{True Parameter$^{\mathrm{c}}$}} & \multicolumn{2}{l|}{\multirow{4}{*}{Estimate}} & \multicolumn{8}{c|}{Model Type}\\ \cline{6-13} \multicolumn{3}{|l|}{}  & \multicolumn{2}{l|}{} & \multicolumn{4}{c|}{SCAD penalty function} & \multicolumn{4}{c|}{No Penalty function}\\ \cline{6-13}   
				\multicolumn{3}{|l|}{}  & \multicolumn{2}{l|}{} & \multicolumn{4}{c|}{Sample size} & \multicolumn{4}{c|}{Sample size}  \\ \cline{6-13} 
				
				\multicolumn{3}{|l|}{}  & \multicolumn{2}{l|}{}  & \multicolumn{1}{c|}{2000} & \multicolumn{1}{c|}{5000} & \multicolumn{1}{c|}{8000} & \multicolumn{1}{c|}{10000} & \multicolumn{1}{c|}{2000} & \multicolumn{1}{c|}{5000} & \multicolumn{1}{c|}{8000} & \multicolumn{1}{c|}{10000} \\ \hline  \hline
				
				\multicolumn{3}{|c|}{\multirow{3}{*}{$\theta_1$=2}}      & \multicolumn{2}{c|}{Mean} & 2.0061 & 2.0049 & 2.0043 & 2.0040 & 2.0339 & 2.0333 & 2.0294 & 2.0215    \\ \cline{4-13} \multicolumn{3}{|c|}{} & \multicolumn{2}{c|}{Median}& 2.0057 & 2.0043 & 2.0034 & 2.0033 & 2.0339 &  2.0336 & 2.0296 & 2.0207    \\ \cline{4-13} \multicolumn{3}{|c|}{} & \multicolumn{2}{c|}{Std} & 0.2434 & 0.1542 & 0.1212 & 0.1079 & 0.2288 & 0.1468 & 0.1162 & 0.1037 \\ \hline \hline
				
				\multicolumn{3}{|c|}{\multirow{3}{*}{$\theta_2$=4}} & \multicolumn{2}{c|}{Mean}  & 3.9687 & 4.0037 & 4.0026  & 3.9984 & 3.9518 & 3.9685 & 3.9701 & 3.9729    \\ \cline{4-13} \multicolumn{3}{|c|}{} & \multicolumn{2}{c|}{Median} & 	3.9797 & 4.0068  & 3.9957 & 4.0016 & 3.9595 & 3.9684  & 3.9692 & 3.9762    \\ \cline{4-13}  \multicolumn{3}{|c|}{} & \multicolumn{2}{c|}{Std}&  1.0359 & 0.6438 & 0.5065 & 0.4544 & 0.9912 & 0.6161 & 0.4847 & 0.4374  \\ \hline
				
	\end{tabular}\end{adjustbox} }\end{center}
	\scriptsize\begin{minipage}{8.2cm}%
		$\myspace$ \\ \textbf{Note:} $^{\mathrm{a}}$Nonlinear Least Squares.\\  $^{\mathrm{b}}$ We estimate the model given a truncated data set using a refined Vox-Populi specification that consists of a sequence of estimated participants' shares, each obtained by employing latent classes analysis, given a specific number of latent classes (as depicted in \eqref{Penalized:regression}). The goal of the penalty function is to find the proper number of latent classes that best fits the data set. The first model specification utilizes the SCAD penalty function and is estimated by a semiparametric penalized non-linear least squares (NLS) method, while the second model specification is estimated without employing a penalty function by a conventional semiparametric non-linear least squares (NLS) method. The bias term function, the nonlinear part of the regression, is approximated using a transformed Fourier series as depicted in section \ref{Sec:Esitmate:Fourier}. Then, we calculate for these estimates the mean, median and standard deviation (Std.) over all data sets. The standard deviations are obtained using the estimates from the Monte Carlo simulations.\\ $^{\mathrm{c}}$ The parameters that are used in the data generation process. $\theta_1$ and $\theta_2$ are the parameters of interest in the substantive equation.
	\end{minipage}		
	\footnotesize
	\renewcommand{\baselineskip}{11pt}	 
\end{table}

Entries presented in Table \ref{tabC} indicate that the substantive equation's estimated parameters obtained in the refined Vox Populi model, using a SCAD penalty function (in the absence of a prior knowledge regarding the number of latent classes). These estimated parameters, $\widehat{\theta}_1=2.0049$ and $\widehat{\theta}_2=4.0037$, almost mimic the parameters estimates that would have been generated from a non-truncated sample, which are $\widehat{\theta}_1=2.000$ and $\widehat{\theta}_2=4.000$, respectively, for a sample size of 5,000 observations. It is worth noting that the parameters estimates' accuracy is improved by employing the SCAD penalty function (in terms of proximity to the true parameters values) relative to the parameters' estimates obtained in the absence of a penalty function, which are $\widehat{\theta}_1=2.0333$ and $\widehat{\theta}_2=3.9518$, using the same sample size. For a given sample size, the estimated standard deviations are slightly smaller in the model without a penalty function relative to the model with the SCAD penalty function. This implies that the penalty function reduces the bias in the estimates at the cost of a minor increase in dispersion.

As reflected by entries in the above tables, correcting for endogenous truncation bias is accurately achieved by applying our semiparametric Sieve estimator, which embeds the notion of refined Vox-Populi decision making.

\section{Conclusion}\label{sec:Summary}
\raggedbottom
The primary purpose of this paper is to correct for selectivity bias generated by endogenous truncation. Incorporating behavioral aspects from economics, psychology and management science to introduce cognition into the participation decision-making allows for endogeneity to take place, to be modeled and to be controlled. We treat each data point's truncation decision based on the decision made by its reference group's opinion space. To accomplish this, we refine the monolithic notion of \textit{vox populi} (\textit{wisdom of the crowd}) by treating the data as a mixture of reference groups.  
We offer a three-stage procedure to correct for this endogenous selectivity bias. In the first stage, latent classes analysis is employed to estimate the various reference groups' memberships based on results from an auxiliary survey data.  In the second stage, estimates for the groups' participation decisions are obtained by averaging their respective group members opinions. 
In the third stage, a semiparametric truncated sample selection model is estimated, consisting of a substantive equation and a selection equation, in which the estimated group's participation decision is an additional covariate. 

The number of reference groups is not arbitrarily imposed but rather estimated using the smoothly clipped absolute deviation (SCAD) penalization mechanism. Monte Carlo simulations involve 2,000,000 different distribution functions, which are not restricted to the unimodal symmetric family of distribution function.  This practically generates 100 million realizations which are not i.i.d. They attest to  a very high accuracy of the model, as depicted by the parameter estimates, which quite accurately mimic the true parameters.
\appendices
\section{Latent classes model's assumptions}
In this section we impose the latent classes model assumptions.

Homogeneity (H) The core assumption in latent class analysis is that the population consists of a set
of mutually exclusive and homogeneous subgroups called classes. The individuals
within a sub-group are homogeneous in the sense that the probability for a particular
response on a particular item depends only on the latent class to which the
individual belongs.

\begin{IEEEeqnarray}{lCr}
Pr(\mathlcal{y}_j = k_j|\Psi_{\mathlcal{k}}=t) = \pi_{jkt}
\end{IEEEeqnarray}

Local Independence (LI)
Local independence assumes that the observed manifest variables, $\mathlcal{y}_1,...,\mathlcal{y}_j$ are related only due to the latent class $\Psi_{\mathlcal{k}}$. Under this assumption, the joint probability of $\mathlcal{y}_1,...,\mathlcal{y}_j$
given $\Psi_{\mathlcal{k}}$ can be written as the product of probabilities of $Y_t$ given the latent class $t$.

Unidimensionality (U)
The assumption of unidimensionality posits that the observed categorical variables
Y are assumed to measure only one ability, attitude, trait, or attribute.

Monotonicity (M)
To obtain stochastic ordering among the latent classes within an item, Croon (1991)
proposed an ordinal latent class model by imposing inequality restrictions:
\begin{IEEEeqnarray}{lCr}
\mathrm{Pr}(\mathlcal{y}_j\le k|\Psi_{\mathlcal{k}}=t_1)\ge \mathrm{Pr}(\mathrm{\mathlcal{y}_j}\le k|\Psi_{\mathlcal{k}}=t_2)\\
\hspace{2em}\pi_{jkt_1}\ge \pi_{jkt_2}\nonumber
\end{IEEEeqnarray}
for all $j$ and $k$, and for all $t_1$ and $t_2$ such that $t_1<t_2$.

\newcommand\nabf[1]{\nabla\Bigger{\mathfrak{f}}\close{\B{\varphi}_{\lambda_n}^{{#1}}}}
\newcommand\frakf[1]{\mathlarger{\mathfrak{f}}\close{\B{\varphi}_{\lambda_n}^{{#1}}}}
\newcommand\betaf[1]{\B{\varphi}_{\lambda_n}^{({#1})}}

\section{The penalty function}\label{Section:Penalty:function:Decomposition:Appendix}
The penalty function $p_{\lambda_n}(\cdotp)$ in \eqref{Problem:convex} is decomposed as $p_{\lambda_n}(\cdotp)=\mathlcal{h}_1(\cdotp)-\mathlcal{h}_2(\cdotp)$ which is a difference of two convex functions: 
\begin{IEEEeqnarray}{lCr}
\mathlcal{h}_1(\cdotp)=\lambda_n\abs{\cdotp},  \hspace{1em} \mathlcal{h}_2(\cdotp)=\lambda_n\abs{\cdotp}-p_{\lambda_n}(\cdotp)
\end{IEEEeqnarray}

Let $b^{(t)}$ be a parameter value obtained at iteration $t$. The best affine approximation of $\mathlcal{h}_2$ at $b^{(t)}$ is:
\begin{IEEEeqnarray}{lCr}\label{Affine:Approximation}
\mathlcal{h}_2(b^{(t+1)})\approx  \tilde{\tilde{\mathlcal{h}}}_2({b^{(t+1)})=\mathlcal{h}_2(b^{(t)})+(b^{(t+1)}-b^{(t)}) \frac{\partial}{\partial b^{(t)}} \mathlcal{h}_2(b^{(t)}}) \nonumber
\end{IEEEeqnarray}

The penalty function is approximated using \eqref{Affine:Approximation} as: 
\begin{IEEEeqnarray}{lCr}
p_{\lambda_n}(b^{(t+1)})\approx \tilde{\tilde{ p}}_{\lambda_n}(b^{(t+1)})=\mathlcal{h}_1(b^{(t+1)})-\tilde{\tilde{\mathlcal{h}}}_2(b^{(t+1)}) \nonumber
\end{IEEEeqnarray}

It is worth noting the following equivalence which must be satisfied: 
\begin{IEEEeqnarray}{lCr}
\underset{b^{(t+1)}}{\arg \min} \hspace{1em}\tilde{\tilde{ p}}_{\lambda_n}(b^{(t+1)})\\ = \underset{b^{(t+1)}}{\arg \min}\hspace{1em}  \mathlcal{h}_1(b^{(t+1)})-(b^{(t+1)})\frac{\partial}{\partial b^{(t)}}  \mathlcal{h}_2(b^{(t)})  \nonumber
\end{IEEEeqnarray}

The algorithm is to solve iteratively the problem using the affine approximation in \eqref{Affine:Approximation}:
\begin{IEEEeqnarray}{lCr}\label{DC:Program}
\B{\varphi}_{\lambda_n^*}^{(t+1)}=\underset{\B{\varphi}_{\lambda_n^*}}{\arg \min}  \LeftCurly{2.5}{\frac{1}{2}\normsq{\B{y_{_1}}-\mathcal{F}(\B{\varphi}_{\lambda_n^*})}{2}}\\+\RightCurly{2}{n\sum_{j=1}^{J}\mathlcal{h}_1(\beta_{j_{\lambda_n}}^{+}+\beta_{j_{\lambda_n}}^{-})- (\beta_{j_{\lambda_n}}^{+}+\beta_{j_{\lambda_n}}^{-})\frac{\partial}{\partial b^{(t)}}  \mathlcal{h}_2(b^{(t)}) }\nonumber
\end{IEEEeqnarray}
\begin{IEEEeqnarray}{lCr}
\text{s.t.}\hspace{1em} \B{\beta_{\lambda_n}^{+}}\ge \B{0}, \hspace{1em} \B{\beta_{\lambda_n}^{-}}\ge \B{0} \nonumber
\end{IEEEeqnarray}
where $t$ is the iteration number and $\nabla \mathlcal{h}_2$ is the gradient of $\mathlcal{h}_2$ evaluated at $\B{\beta_{\lambda_n}^{+,(t)}}+\B{\beta_{\lambda_n}^{-,(t)}}$, such that $\close{\B{\beta_{\lambda_n}^{+,(t)}},\B{\beta_{\lambda_n}^{-,(t)}}}\in \B{\varphi}_{\lambda_n^*}^{(t)}$.

After decomposing the coefficient vector $\B{\beta}$ to enable difference of convex functions (DC) programming formulation, problem \eqref{DC:Program} can be formulated as a weighted LASSO problem:
\begin{IEEEeqnarray}{lCr}\label{Weighted:LASSO}
\\ \B{\varphi}_{\lambda_n}^{(t+1)}=\underset{\B{\varphi}_{\lambda_n}}{\arg \min}  \curly{\frac{1}{2}\normsq{\B{y_{1}-\mathcal{F}(\B{\varphi}_{\lambda_n^*})}}{2}+\sum_{j=1}^{J}\tilde\lambda_{j}\abs{\beta_{j_{\lambda_n}}}}\nonumber
\end{IEEEeqnarray}
where $\B{\tilde\lambda}=n(\lambda_{n}-\nabla \mathlcal{h}_2)$ and its $j$th element is $\tilde\lambda_{j}$.

Let $\Bigger{\mathfrak{f}}(\B{x})\equiv \frac{1}{2}\normsq{\B{y_{_1}}-\mathcal{F}(\B{x})}{2}$, the update rule to minimize \eqref{Weighted:LASSO} is computed using a second-order approximation of $\Bigger{\mathfrak{f}}(.)$ at $\B{\varphi}_{\lambda_n}^{(k)}$ \cite{yang2013review}:
\begin{IEEEeqnarray}{lCr}\label{Gradient:Proximal:Hess}
\\ \B{\varphi}_{\lambda_n}^{(k+1)}=\underset{\B{\varphi}_{\lambda_n}}{\arg \min}\Bigg\{\frakf{k}+(\B{\varphi}_{\lambda_n}-\B{\varphi}_{\lambda_n}^{(k)})^T\nabla \frakf{k}\nonumber
\end{IEEEeqnarray}
\begin{IEEEeqnarray}{lCr}
\hspace{4em}+\frac{1}{2}(\B{\varphi}_{\lambda_n}-\B{\varphi}_{\lambda_n}^{(k)})^T\nabla^2\frakf{k}(\B{\varphi}_{\lambda_n}-\B{\varphi}_{\lambda_n}^{(k)})\nonumber\\\hspace{4em}+\sum_{j=1}^{J}\tilde\lambda_{j}\abs{\beta_{j_{\lambda_n}}}\Bigg\}\nonumber
\end{IEEEeqnarray}

To simplify the minimization problem in \eqref{Gradient:Proximal:Hess} we let $\alpha I\cong\nabla^2\mathfrak{f}(\B{\varphi}_{\lambda_n}^{({k})})$, the following approximation is used:
\begin{IEEEeqnarray}{lCr} 
\alpha=\frac{\close{\nabf{k}-\nabf{k-1}}^T\close{\betaf{k}-\betaf{k-1}}}{\close{\betaf{k}-\betaf{k-1}}^T\close{\betaf{k}-\betaf{k-1}}}
\end{IEEEeqnarray}
where $I$ is the identity matrix.

to get:
\begin{IEEEeqnarray}{lCr}\label{Proximal:Gradient:approximated}
\B{\varphi}_{\lambda_n}^{(k+1)}=\underset{\B{\varphi}_{\lambda_n}}{\arg \min}\Bigg\{\frakf{k}+(\B{\varphi}_{\lambda_n}-\B{\varphi}_{\lambda_n}^{(k)})^T\nabla\\ \frakf{k}+\frac{\alpha}{2}\normsq{\B{\varphi}_{\lambda_n}-\B{\varphi}_{\lambda_n}^{(k)}}{2}+\sum_{j=1}^{J}\tilde\lambda_{j}\abs{\beta_{j_{\lambda_n}}}\Bigg\}\nonumber
\end{IEEEeqnarray}

After canceling out the constant terms in \eqref{Proximal:Gradient:approximated} (which are not function of $\B{\varphi}_{\lambda_n}$), the minimization problem is reformulated as follows:
\begin{IEEEeqnarray}{lCr}
\\ \B{\varphi}_{\lambda_n}^{(k+1)}=\underset{\B{\varphi}_{\lambda_n}}{\arg \min}\Bigg\{\frac{1}{2}\normsq{\B{\varphi}_{\lambda_n}-\B{u^{(t)}}}{2}+\frac{1}{\alpha}\sum_{j=1}^{J}\tilde\lambda_{j}\abs{\beta_{j_{\lambda_n}}}\Bigg\}\nonumber
\end{IEEEeqnarray}
where $\B{u^{(t)}}=\B{\varphi}_{\lambda_n}^{(t)}-\frac{1}{\alpha_t}\nabla \mathcal{F}(\B{\varphi}_{\lambda_n}^{(t)})$ and $u_j^{(t)}$ is its $j$'th element.

The algorithm for solving \eqref{Weighted:LASSO} is to update each $j$ component in the parameter vector $\B{\varphi}_{\lambda_n}$ using the well-known soft-threshold algorithm (\cite{yang2013review} and \cite{yang2016sparse} for linear and non-linear treatments respectively):\footnote{Since the $\ell_1$ norm (the weighted LASSO penalty function) is separable, the computation of $\B{\varphi}_{\lambda_n}^{(k+1)}$ reduced to solve a one dimensional minimization problem for each of its components.}
\begin{IEEEeqnarray}{lCr}
\varphi_{j_{\lambda_n}}^{(t+1)}=\text{soft}(u_j^{(t)},\frac{\tilde\lambda_j}{\alpha_t})
\end{IEEEeqnarray}
where $\text{soft}(u,a)\equiv \mathrm{sign}(u)\max(\abs{u}-a,0)$.

For any parameter $j$ which is not part of the penalization $\tilde\lambda_j=0$. 
\section{The modified Soft thresholding algorithm}
Let define $p_{\lambda_n}^{\prime}(\cdot)$ as:
\begin{IEEEeqnarray}{lCr}
p_{\lambda_n}^{\prime}(v)=\begin{cases}
\lambda_n \mathrm{sign}(v) & \text{if } \abs{v}\le\lambda_n\\
\fracBig{a\lambda_n \mathrm{sign}(v)-v}{(a-1)} & \text{if } \lambda_n<\abs{v}\le a\lambda_n\\
0 &\text{if } \abs{v}> a\lambda_n
\end{cases}
\end{IEEEeqnarray}
where $\mathrm{sign}(v)=1\curly{v>0}-1\curly{v<0}$ such that $1\curly{\cdot}$ is an indicator function. 

We characterize the second order approximation for the non-linear function in (.):
\begin{IEEEeqnarray}{lCr}
\B{\beta}_{\lambda_n}^{(t+1)}=\underset{\B{\beta}}{\arg\max}\hspace{1em} (\B{\beta}-\B{\beta}_{\lambda_n}^{(t)})^T\nabla \mathfrak{f}(\B{\beta}_{\lambda_n}^{(t)})\\+\frac{1}{2}(\B{\beta}-\B{\beta}_{\lambda_n}^{(t)})^T\nabla^2 \mathfrak{f}(\B{\beta}_{\lambda_n}^{(t)})(\B{\beta}-\B{\beta}_{\lambda_n}^{(t)})+\sum_{k=1}^{\tilde{G}_{\max}}p_{\lambda_n}\close{\abs{\beta_{k_{\lambda_n}}^{(t+1)}}}\nonumber
\end{IEEEeqnarray}

Taking the derivative with respect to $\B{\beta}$ to obtain:
\begin{IEEEeqnarray}{lCr}\label{Approximate:penalized:F.o.c}
\nabla \mathfrak{f}(\B{\beta}_{\lambda_n}^{(t)})+\nabla^2 \mathfrak{f}(\B{\beta}_{\lambda_n}^{(t)})(\B{\beta}^{(t+1)}-\B{\beta}_{\lambda_n}^{(t)})\\+\nabla\close{\sum_{k=1}^{\tilde{G}_{\max}}p_{\lambda_n}\close{\abs{\beta_{k_{\lambda_n}}^{(t+1)}}}}=0\nonumber
\end{IEEEeqnarray}
where $\nabla\close{\sum_{k=1}^{\tilde{G}_{\max}}p_{\lambda_n}\close{\abs{\beta_{k_{\lambda_n}}^{(t+1)}}}}$ is a vector of size $\tilde{G}_{\max}\times 1$ and its $k$'th element is $p_{\lambda_n}^{\prime}(\beta_{k_{\lambda_n}}^{(t+1)})$.

Let $\mathcal{A}=\LeftCurly{1.8}{\braket{\mathfrak{s}_1,...,\mathfrak{s}_{\tilde{G}_{\max}}}^T| \hspace{1em}\mathfrak{s}_{k}\in\curly{-1,0,1}},\\ \RightCurly{1.8}{\hspace{1em} k=1,...,\tilde{G}_{\max}}$ be the set consisting of all possible signs for a real number vector of size $\tilde{G}_{\max}\times 1$. Using this set,  we denote a sign operator $\mathfrak{S}:\mathbb{R}^{\tilde{G}_{\max}}\mapsto \mathcal{A}$. It follows that the $k$'th element of $\mathfrak{S}(\B{\beta}^{(t+1)})$ is $\mathrm{sign}(\beta_k^{(t+1)})$ and the matrix representation of
$\nabla\close{\sum_{k=1}^{\tilde{G}_{\max}}p_{\lambda_n}\close{\abs{\beta_{k_{\lambda_n}}^{(t+1)}}}}$ is:\footnote{The notation $\circ$ represents the Hadamard product.}
\begin{IEEEeqnarray}{lCr}\label{Deriv:Penalty:Gradient}
\nabla\close{\sum_{k=1}^{\tilde{G}_{\max}}p_{\lambda_n}\close{\abs{\beta_{k_{\lambda_n}}^{(t+1)}}}}=\lambda_n\B{\mathlcal{r}}\circ\mathfrak{S}(\B{\beta}^{(t+1)})- \B{\mathcal{R}}\B{\beta}^{(t+1)}\nonumber
\end{IEEEeqnarray}
where $\B{\mathcal{R}}$ and $\B{\mathlcal{r}}$ are a matrix of size $\tilde{G}_{\max}\times \tilde{G}_{\max}$  and a $\tilde{G}_{\max}\times 1$  vector, defined in \eqref{R:matrix} and \eqref{r:vector}, respectively.

\begin{IEEEeqnarray}{lCr}\label{R:matrix}
\B{\mathcal{R}}\equiv \frac{1}{a-1}\\ \times\braket{
	\begin{matrix}
	1\curly{\lambda_n < \abs{\beta_1}\le a\lambda_n} &  \cdots & 0 \\ 
	\colon & \ddots & \colon \\
	0 & \cdots & 1\curly{\lambda_n < \abs{\beta_{\tilde{G}_{\max}}}\le a\lambda_n} \nonumber
	\end{matrix}}
\end{IEEEeqnarray}

\begin{IEEEeqnarray}{lCr}\label{r:vector}
\B{\mathlcal{r}}\equiv\braket{
	\begin{matrix}
	1\curly{\abs{\beta_1}\le \lambda_n}+\fracBig{a}{a-1}1\curly{\lambda_n < \abs{\beta_1}\le a\lambda_n}\\ 
	\colon\\
	1\curly{\abs{\beta_{\tilde{G}_{\max}}}\le \lambda_n}+\fracBig{a}{a-1}1\curly{\lambda_n < \abs{\beta_{\tilde{G}_{\max}}}\le a\lambda_n}
	\end{matrix}}\nonumber
\end{IEEEeqnarray}

The matrix representation depicted in \eqref{Deriv:Penalty:Gradient} implies that $\B{\beta}$ can be isolated from \eqref{Approximate:penalized:F.o.c} to get:
\begin{IEEEeqnarray}{lCr}\label{b:sign:equ}
\B{\beta}_{\lambda_n}^{(t+1)}=\B{\beta}_{\lambda_n}^{(t)}-\braket{\nabla^2 \mathfrak{f}(\B{\beta}_{\lambda_n}^{(t)})-\B{\mathcal{R}}}^{-1}\\\times\close{\nabla\mathfrak{f}(\B{\beta}_{\lambda_n}^{(t)})-\lambda_n\B{\mathlcal{r}}\circ \mathfrak{S}(\B{\beta}_{\lambda_n}^{(t+1)})}\nonumber
\end{IEEEeqnarray}
However, the expression in \eqref{b:sign:equ} depends on the sign operator which is a function of $\B{\beta}_{\lambda_n}^{(t+1)}$ to alleviate the recursive nature of the formula for $\B{\beta}_{\lambda_n}^{(t+1)}$  we substitute $\mathfrak{S}(\B{\beta}^{(t+1)})$ with a vector of signs  $\mathfrak{g}\in\mathcal{A}$. It worth noting that  $\mathfrak{g}=\mathfrak{S}(\B{\beta}_{\lambda_n}^{(t+1)})$  if and only if: 
\begin{IEEEeqnarray}{lCr}\label{Find:Signs}
\mathfrak{S}\close{\B{\beta}_{\lambda_n}^{(t)}-\braket{\nabla^2 \mathfrak{f}(\B{\beta}_{\lambda_n}^{(t)})-\B{\mathcal{R}}}^{-1}\close{\nabla\mathfrak{f}(\B{\beta}_{\lambda_n}^{(t)})-\lambda_n\B{\mathlcal{r}}\circ \mathfrak{g}}}=\mathfrak{g}\nonumber
\end{IEEEeqnarray}

In cases where a solution $\mathfrak{g}$ to \eqref{Find:Signs} is found, the updated $\B{\beta}$ is:
\begin{IEEEeqnarray}{lCr}
\B{\beta}_{\lambda_n}^{(t+1)}=\B{\beta}_{\lambda_n}^{(t)}-\braket{\nabla^2 \mathfrak{f}(\B{\beta}_{\lambda_n}^{(t)})-\B{\mathcal{R}}}^{-1}\close{\nabla\mathfrak{f}(\B{\beta}_{\lambda_n}^{(t)})-\lambda_n\B{\mathlcal{r}}\circ \mathfrak{g}}\nonumber
\end{IEEEeqnarray} 

The justification for the proposed non-linear penalized regression estimation algorithm is based on a unified algorithm introduced by \cite{fan2001variable} which optimizes various linear penalized regression problems via local quadratic approximations.

In cases where there is no solution, we find the largest subset of signs for which there is a solution to \eqref{Find:Signs}, and set to zero all the rest of the parameter values as if we employed the original soft-thresholding algorithm.

\section{Binary Response model assumptions}\label{Sec:Binary:Assumptions}

Our objective here is present the necessary conditions for identification of the binary response model unknown parameters. These necessary conditions are depicted in the following assumptions:\footnote{The assumptions are taken from \cite{brock2007identification}.}
\begin{assumption}\label{Assumption:IID} (i.i.d and symmetry)
	Conditional on $(\B{z},\B{x}_g,\psi_g)$, the random payoff terms $\disturb_{2i}$ are independently and	identically distributed according to $F_{\mathrm{\disturb_2}}$, such that $F_{\mathrm{\disturb_2}}(0)=0.5$.\footnote{This restriction is imposed for the intercept identification \cite{xu2015maximum}.}
\end{assumption}

\begin{assumption} (continuity)
	$F_{\mathrm{\disturb_2}}$ is absolutely continuous with associated density $dF_{\mathrm{\disturb_2}}$; $dF_{\mathrm{\disturb_2}}$ is positive almost
	everywhere on the support $(L,U)$ which may be $(-\infty,\infty)$.
\end{assumption}

\begin{assumption} (linear independence among the observable
	individual-specific and group-specific characteristics, and variation in $\B{\mathrm{z}}$)
	$\B{z}$ does not include a constant; there exists a group $g_0$ such that $\mathrm{supp}(\B{\mathrm{z}}_{-j|g_0})$\footnote{The support of a random variable $\mathrm{v}$ is denoted by $\mathrm{supp}(\mathrm{v})$. For a vector such as $\B{\mathrm{z}}$, $\B{\mathrm{z}}_{-l}$ denotes the vector when $\mathrm{z}_l$ is omitted.} is not
	contained in a proper linear subspace of $\mathbb{R}^{L_1}$; there exists an $z_l$ (with associated non-zero coefficient $c_l$), such that for almost every value of the vector $\B{\mathrm{z}_{-l|g_0}},\mathrm{x_{1|g_0},...,x_{s|g_0}}$, the conditional
	distribution function of $\mathrm{z}_{l|g_0}$ given $\B{\mathrm{z}_{-l|g_0}},\mathrm{x_{1|g_0},...,x_{s|g_0}}$
	has everywhere positive density.
\end{assumption}

\begin{assumption} (an unbounded support
	assumption on one element of $\B{\mathrm{z_i}}$ and all elements of $\B{\mathrm{x_g}}$)
	the conditional distribution function of $\B{\mathrm{x_g}}$ has everywhere positive density.
\end{assumption}

\begin{assumption}\label{Assumption:Unobserved} (local no unobservable group's characteristics)
	$\alpha_g=0$, $\forall g$
\end{assumption}

Based on Proposition 1 in \cite{brock2007identification}, under assumptions \eqref{Assumption:IID}-\eqref{Assumption:Unobserved}, the parameters of the binary choice model
are identified up to scale.

The contextual effect identification:
\begin{assumption} (exclusion restriction with respect to groups characteristics)
	There must be at least one element in $\B{x_g}$ that does not included in $\B{x_g^c}$.
\end{assumption}

\begin{assumption} (latent classes with non-random assignment)
	The latent classes capture all the unobserved (within group) heterogeneity.	
\end{assumption}

\begin{assumption}\label{Assumption:LocalCovShift} (local covariate shift)
	The conditional distribution function of $\mathrm{y_j}$, $j=1,2$ given the observed covariates and the unobserved heterogeneity (captured by the classes) is same for all the observations characterized with these covariates.
\end{assumption}

The assumption \eqref{Assumption:LocalCovShift} implies that the objective function can change across reference groups, but not within reference group. It also implies Homogeneity within reference group.

There are two core assumptions in latent classes analysis:
Assumption 1: (Homogeneity) The population consists of  a set of mutually exclusive homogeneous subgroups.

Assumption 2: (Local Independence) The vector of observed characteristics, $[\mathlcal{y}_1,...,\mathlcal{y}_J]^T$ are related only due to the latent classes.\footnote{Latent classes can be modeled nonparametrically. However, for computational convenience, we model the outcome variables, determined by the latent classes (labels), parametrically using an ordinal logistic regression for each outcome variable.}

\section*{Acknowledgment}
We thank Larry Manevitz for very constructive comments and Omiros Papaspiliopoulos for very constructive conversations. 

\bibliographystyle{ieeetran}
\bibliography{reference2}

\begin{IEEEbiography}[{\includegraphics[width=2in,height=1in,clip,keepaspectratio]{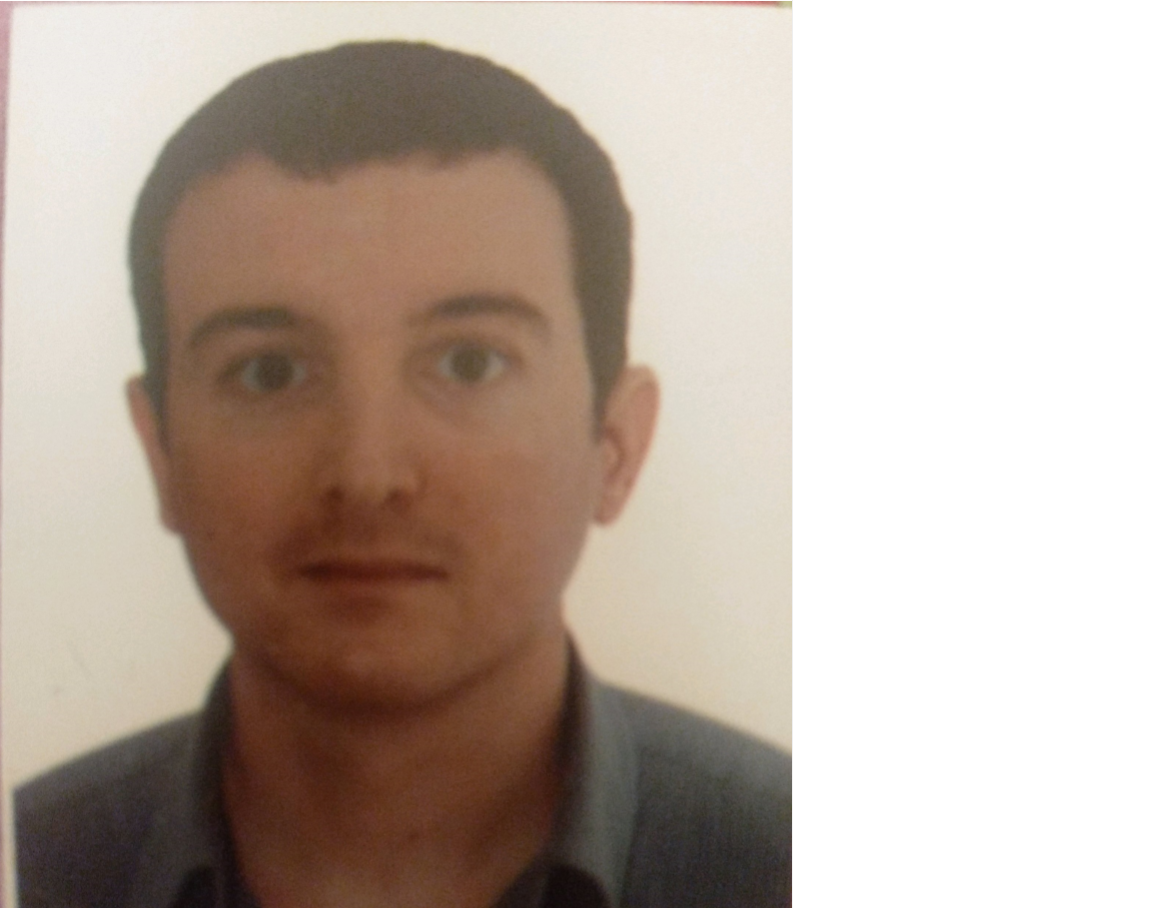}}]{Nir Billfeld} is a researcher at the university of Haifa, Israel. He received the B.A. in economics and statistics from the university of Haifa (2006), Israel, M.A. in economics from Tel-Aviv university (2010). Currently Ph.D. the university of Haifa (2018). 
\end{IEEEbiography}

\begin{IEEEbiography}[{\includegraphics[width=1in,height=1.25in,clip,keepaspectratio]{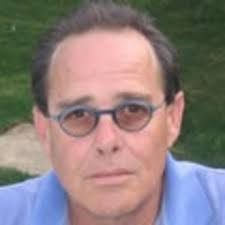}}]{Moshe Kim} is professor of economics at the University of Haifa, Israel. He is the founder and former director of Barcelona Banking Summer School at the Universitat Pompeu Fabra, Barcelona, former director of the endowed chair of banking at Humboldt University of Berlin, Senior Distinguished Fellow at the Swedish School of Economics in Helsinki (Hanken), institute research professor at the German Institute for Economic Research (DIW), consultant at the Central Bank of Norway and recently visited NYU Shanghai. He was recently declared high end foreign expert by the Chinese foreign ministry and is a recent recipient of the Outstanding Tutor Award from the Chinese Ministry of Education. He holds a PhD from the University of Toronto. Kim's research interests are econometrics, banking, financial markets, and industrial organization. His books include Microeconometrics of Banking: Methods, Applications, and Results (Oxford University Press, 2009). His work has appeared in the Journal of Finance, the Journal of Monetary Economics, the Journal of Financial Intermediation, the Journal of Business and Economics Statistics, the Journal of Money Credit and Banking, International Economic Review, the Journal of Accounting and Economics, the Journal of Public Economics, the Journal of Urban Economics, the Journal of Law and Economics, the Journal of Banking and Finance, the Journal of Industrial Economics, the International Journal of Industrial organization. 
\end{IEEEbiography}

\EOD

\end{document}